\documentclass{aa}

\usepackage[utf8]{inputenc}
\usepackage{natbib,twoopt}
\usepackage{graphicx}
\usepackage{booktabs,caption}
\usepackage{xcolor}
\usepackage[breaklinks=true]{hyperref} 
\bibpunct{(}{)}{;}{a}{}{,}             
\makeatletter
  \newcommandtwoopt{\citeads}[3][][]{\href{http://adsabs.harvard.edu/abs/#3}%
    {\def\hyper@linkstart##1##2{}%
     \let\hyper@linkend\@empty\citealp[#1][#2]{#3}}}
  \newcommandtwoopt{\citepads}[3][][]{\href{http://adsabs.harvard.edu/abs/#3}%
    {\def\hyper@linkstart##1##2{}%
     \let\hyper@linkend\@empty\citep[#1][#2]{#3}}}
  \newcommandtwoopt{\citetads}[3][][]{\href{http://adsabs.harvard.edu/abs/#3}%
    {\def\hyper@linkstart##1##2{}%
     \let\hyper@linkend\@empty\citet[#1][#2]{#3}}}
  \newcommandtwoopt{\citeyearads}[3][][]%
    {\href{http://adsabs.harvard.edu/abs/#3}
    {\def\hyper@linkstart##1##2{}%
     \let\hyper@linkend\@empty\citeyear[#1][#2]{#3}}}
\makeatother

\usepackage{ulem}
\usepackage{mathabx}

\title{Forming planets around stars with non-solar elemental composition}
\titlerunning{Forming planets with non-solar elemental composition}
\author{D.M. Jorge\inst{1}
          ,
          I.E.E. Kamp\inst{1}
          ,
          L.B.F.M. Waters\inst{2,3}
          ,
          P. Woitke\inst{4,5,6}
          ,
          R.J. Spaargaren\inst{7}
          }

   \institute{Kapteyn Astronomical Institute, University of Groningen, PO Box 800, 9700AV Groningen, The Netherlands\\
              \email{jorge@astro.rug.nl}
         \and
         Department of Astrophysics/IMAPP, Radboud University, PO Box 9010, 6500 GL Nijmegen, the Netherlands
         \and
         SRON Netherlands Institute for Space Research, Niels Bohrweg 4, 2333 CA Leiden, the Netherlands
         \and
         Space Research Institute, Austrian Academy of Sciences, Schmiedlstrasse 6, A-8042 Graz, Austria
         \and
         SUPA, School of Physics \& Astronomy, St Andrews University, North Haugh, St Andrews, KY169SS, UK
         \and
         Centre for Exoplanet Science, University of St Andrews, North Haugh, St Andrews, KY169SS, UK
         \and
         Institute of Geophysics, ETH Zurich, Sonneggstrasse 5, 8092, Zurich, Switzerland
             }
\authorrunning{Jorge et al.}
\date{Received / Accepted}

\abstract{
Stars in the solar neighbourhood have refractory element ratios slightly different from the Sun. It is unclear how much the condensation of solids and thus the composition of planets forming around these stars is affected.}
{We aim to understand the impact of changing the ratios of refractory elements Mg, Si, and Fe within the range observed in solar type stars within 150~pc on the composition of planets forming around them.}
{We use the GGchem code to simulate the condensation of solids in protoplanetary disks with a Minimum Mass Solar Nebula around main sequence G-type stars in the Solar neighbourhood. We extract the stellar elemental composition from the Hypatia database.}
{We find that a lower Mg/Si ratio shifts the condensation sequence from forsterite (Mg$_2$SiO$_4$) and SiO to enstatite (MgSiO$_3$) and quartz (SiO$_2$); a lower Fe/S ratio leads to the formation of FeS and FeS$_2$ and little or no Fe-bearing silicates. Ratios of refractory elements translate directly from the gas phase to the condensed phase for $T\,<\,1000$~K. However, ratios with respect to volatile elements (e.g.\ oxygen and sulphur) in the condensates --- the building blocks of planets --- differ from the original stellar composition.}
{Our study shows that the composition of planets crucially depends on the abundances of the stellar system under investigation. Our results can have important implications for planet interiors, which depend strongly on the degree of oxidation and the sulphur abundance.}


\begin{document}

\maketitle

\section{Introduction}
\label{sec:intro}


In the interstellar medium, molecular clouds give rise to the conditions necessary for star formation.  The formation of a new star proceeds through a flat rotating accretion disk. Hence, the primordial composition of the disk and the host star share the same elemental abundances as the one in the collapsing cloud.  In the innermost disk (inside a few~au), the material can undergo several cycles of condensation and evaporation of dust due to various processes such as episodic accretion \citep{Min2011, audard2014}. Subsequently, these dust particles agglomerate forming solid bodies increasing in size and mass, and eventually turn into a planet orbiting its host star \citep[see][for a recent review]{raymond2020}. 

The physical and chemical processes happening during this formation are influenced by many factors, including the temperature, the pressure, and chemical composition of the planet forming disk during the condensation. In fact \citet{adibekyan2021chemical} confirm the close connection between stellar and rocky planetary iron composition; however, they  also note that additional processes during planet formation can lead to planets incorporating less iron than available \citep{2019ESS.....430006D}. Terrestrial planets are mainly composed of elements such as Mg, Si, Fe and O \citep[see][for a recent overview]{mcdonough2021}. Moreover, volatile elements such as sulphur are very interesting when it comes to determine the composition of the core \citep{laurenz2016}. Hence those are the key elements investigated in this paper. The ratios of these different elements control the planets  composition, thus affecting the mass-radius relationship \citep{valencia2007}. In turn, the mass-radius relationship is also often used to infer the planet interior composition \citep[e.g.][]{Putirka2021,adibekyan2021chemical}.

Differences from the Solar elemental values can be attributed to processes such as Type Ia supernovae \citep[e.g.,][]{Hinkel2014}. Even though Mg and Si are both $\alpha$ capture elements, they are produced in different nuclear burning phases (a function of stellar mass) and can vary with respect to each other throughout the galaxy \citep{Adibekyan2015}. Also, chemodynamical simulations of a Milky-Way-type galaxy show that the ratio of certain elements with respect to Fe can vary \citep{Kobayashi2011a} due to supernova feedback and stellar winds. Therefore, variation from the Solar norm of refractory elements is a well established fact. 
The question remaining to answer is whether, what has been considered by stellar astrophysicists as a small fluctuation in the ratio of the refractory elements (factor two or three difference), has a noticeable impact on the composition of a planet condensing from a planet forming disk. 

Recent work showed that pebble drift can have an effect on planetesimal composition. Dust grains will sublimate around rocklines, thereby allowing some of the refractory components to diffuse back into the gas and alter the composition of surrounding planetesimals \citep{Aguichine2020}. Using a simple chemistry prescription, \citet{Schneider2021a,Schneider2021b} showed that pebble drift will impact the composition of the forming planets and their atmospheres. However, we focus here on the initial step of condensation and leave the long term evolution to subsequent studies.

Research in planet formation often assumes that the elemental ratios found in the gas phase directly translate in the condensed phase from which planets form. However, there is a growing interest in challenging these assumptions. \cite{Bitsch2020} study planet formation scenarios interior and exterior to the water ice line and show that sub- and super-solar metallicity has an influence on the composition of the planetary building blocks. Therefore, more work is needed to investigate the effects of observed element gradients found in the Solar neighbourhood on planet formation scenarios.

The aim of this paper is to investigate the equilibrium condensation sequence for various element mixtures using the computer code GGchem \citep{Woitke2018}. In Sect.~\ref{sec:method} we explain the derivation of the condensation sequences using GGchem and our stellar sample selection for the Solar neighbourhood (element abundances). In Sect.~\ref{sec:results}, we vary the elemental abundances in accordance with observational data and study how these variations propagate into the condensates and thus the chemical composition of the disks that surround our selected main sequence G-type stars. Subsequently, we discuss in Sect.~\ref{sec:discussion} what impact these results have on planet formation. 


\begin{figure}
\centering
  \includegraphics[width=8.5cm]{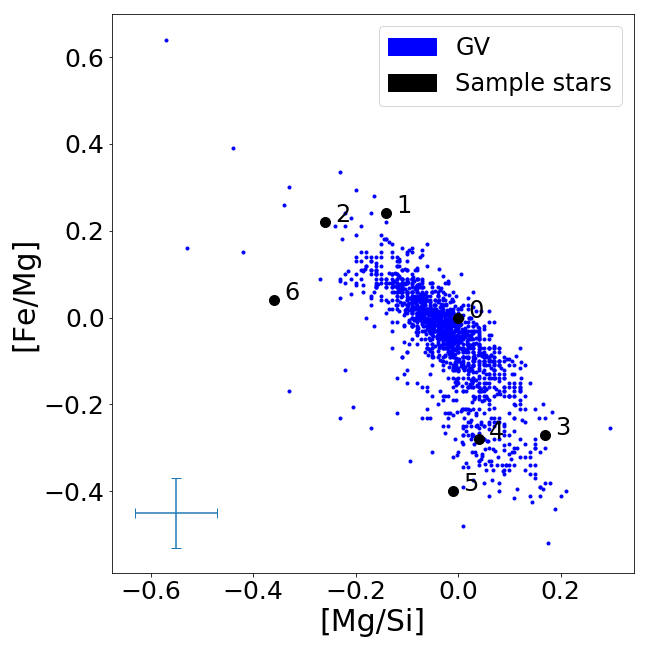}
  \includegraphics[width=8.5cm]{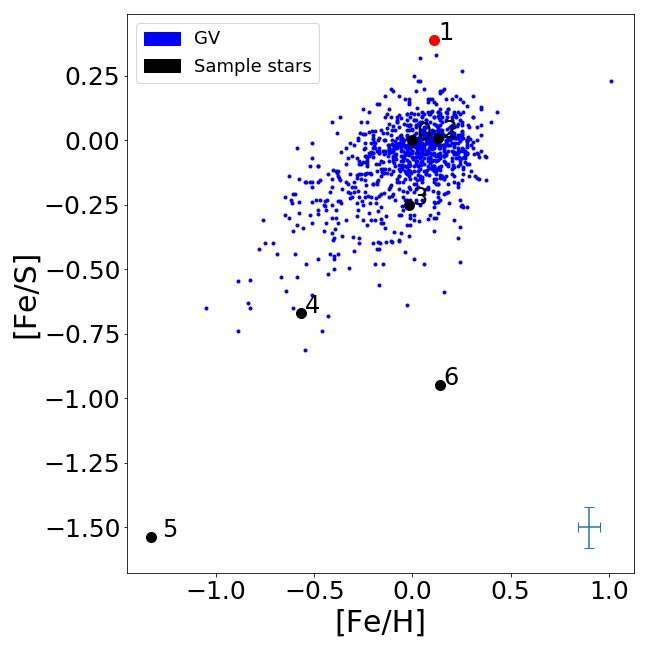}
    \caption{Ratios of the refractory elements Mg, Si and Fe, as well as S, for main sequence G-type stars in the Solar neighbourhood \citep[within 150~pc of the Sun, 1529 stars, from][]{Hinkel2014}. A representative error bar is provided representing the median spread of the respective elements. Black dots represent the location of the sample of stars studied in this paper and described in Table \ref{fig:table sample}. The red dot indicates a star for which the ratio was not available in the Hypatia Database, and which was inferred from the median of the known abundance values (see App.~\ref{sec:chemical}). All values are normalised by the Solar abundance values according to Eq.~\ref{eq:4}.}
    \label{fig:hypatia stars}
\end{figure}

\section{Method}
\label{sec:method}

\subsection{The GGchem code}
\label{sec:code}

GGchem was first developed by \cite{Gail1986} and subsequently improved upon by C.~Dominik, Ch.~Helling and P.~Woitke up until 2005. However, the version currently being used is re-written in a modern FORTRAN-90 code architecture \citep{Woitke2018}. The major improvement that this new version of the code offers is the possibility to go down to a temperature of 100~K, thanks to the use of quadruple precision arithmetic. 

The various thermo-chemical data implemented in the GGchem code are taken from the NIST-JANAF database \citep{Chase1982,Chase1986} and the geophysical SUPCRTBL database \citep{Zimmer2016, Johnson1992}. The GGchem code has been thoroughly benchmarked against the values found in the public thermo-chemical equilibrium abundances (TEA) code \citep{Hinkel2016} and has been found to agree remarkably well.

A total of 24 elements (H, He, Li, C, N, O, Na, Mg, Al, Si, S, Cl, K, Ca, Ti, V, Cr, Mn, Fe, Ni, Zr, W, F, P), as well as charged species, can be selected as input. Furthermore, the initial abundance of each of these chemical elements must be specified. Depending on this initial choice, GGchem will form molecules and condensate species (including water containing minerals such as phyllosilicates and also ices) accordingly. Given a specific temperature and pressure, the GGchem code calculates the various molecules and condensates forming by minimising the Gibbs free energy. 

Because we have no reason to exclude any of the 24 elements available in the GGchem code, we use all of them resulting in a total of 552 molecules and 241 condensates. The abundance of these species is shown in condensation sequences for the given temperature-pressure profile used to model the planet forming disk.

\subsection{Initial elemental abundances}
\label{sec:abund}

\subsubsection{The Hypatia Catalog}
\label{sec:hypatia}

We take, as starting assumption, the host star abundances as proxy for the chemical composition of the original cloud from which both the star and the disk are formed. Recently, \cite{adibekyan2021chemical} confirmed the clear correlation between the iron-mass fraction of a planet and its host star (albeit not a one-to-one correlation).

\begin{table*}[th!]

\caption{Data on the stellar sample taken from the Hypatia database.}
\label{fig:table sample}
\centering
\begin{tabular}{lllllllllll}
\# & HIP    & Fe     & C      & O      & Mg     & Si & S &    SpTy   & d[pc] & $T_{ \text{eff} }$ [K] \\ \hline
0  & Sun    & $7.5\pm 0.04$ & $8.43\pm 0.05$ & $8.69\!\pm\!0.05$ & $7.60\!\pm\!0.04$ & $7.51\!\pm\!0.03$ & $7.12\!\pm\!0.03$ &    G2V    & 0                & 5778             \\ \hline
1  & 109836 & +0.39  & +0.225 & +0.25  & +0.15  & +0.285 & $\textcolor{red}{+0.277}$  & G6V    & 72.8             & 5810             \\
2  & 28869  & +0.13  & +0.01  & +0.02  & -0.09  & +0.17  & +0.12  & G0V    & 47.8             & 6120             \\
3  & 63048  & -0.019 & +0.13  & +0.19  & +0.25  & +0.08  &  +0.23  & G2V    & 43.05            & 5665             \\
4  & 43393  & -0.57  & -0.305 & +0.03  & -0.293 & -0.329 &  +0.1 & G8/K0V & 49.67            & 5306             \\
5  & 99423  & -1.335 & -0.49  & -0.265 & -0.93  & -0.92  &  +0.21 & G0V    & 109.58           & 5745             \\ \hline
6  & 83359  & +0.145 & -0.15  & -0.03  & +0.1   & +0.46  &  +1.1 & G5-K1III     & 110.63           &   4949   
\end{tabular}
\tablefoot{The full set of chemical abundances used for each star can be seen in App.~\ref{sec:chemical}. The red value indicates an elemental abundance that is not available in the Hypatia Database, but is inferred from the median of the known data for the metals of the specific star. Star 0 represents the Sun and is the reference to which any comparison is made. Stars 1-5 are main sequence G-type stars. Star 6 is separated from the rest since it is a Red Giant; its spectral type (SpTy) and effective temperature $T_{\text{eff}}$ are from \cite{Liu2010}.}

\end{table*}

We choose the Hypatia Catalog \citep{Hinkel2014} as the source of our elemental abundances. Its online database comprises a wide variety of measurements of stellar abundances in the Solar neighbourhood (within 150~pc of our Sun) comprising a total (at the time of the simulation) of 6193 FGKM-type stars, gathered from a wide variety of sources (190+). The Hypatia Catalog lists abundances normalised to the Solar abundance values. Throughout our work we use the Solar elemental abundances of \citet{Asplund2009}.

The Hypatia Catalog is a heterogeneous database, as no unique technique has been used in the derivation of the data. For some elements, various sources list different sets of abundances for the same star, sometimes showing a discrepancy from one another larger than the error bar of each individual value. \citet{Hinkel2014} report a mean spread for all elements of 0.14~dex, and a median spread of 0.11~dex. \citet{Blecic2016} investigate these discrepancies between research teams and show that these issues  might be inherent to the methods themselves as much as the protocol used. However, for the purpose of this paper we are only interested in the influence of the typical spread in the ratios of refractory elements (namely Mg, Si and Fe, see Fig.~\ref{fig:hypatia stars} for main sequence G-type stars). 

20$\%$ of the stars show a ratio $|\text{Mg/Si}|>0.1$, while 33$\%$ show a ratio $|\text{Fe/Mg}|>0.1$. The spread in the Fe/S ratio is even more striking. However, only 956 out of the 1529 main sequence G-type stars studied have a sulphur abundance; 43$\%$ of them show a ratio $|\text{Fe/S}|>0.1$.

\subsubsection{The sample of stars}
\label{sec:sample-selection}

To study how variations in elemental abundances affect the condensation of the various elements within the disk, and therefore, the subsequent planet forming bulk, we need to select a sample of stars. We selected from the Hypatia Catalog six stars which are,
\begin{itemize}
    \item main sequence G-type stars
    \item not found in a binary or higher order stellar system.
    \item registered in the Hypatia Catalog with as much chemical abundance data as possible (with a minimum of Mg, Si, Fe, C, and O present).
    \item representative of the typical spreads in the elements ratios that we can expect (see Fig.~\ref{fig:hypatia stars}).
\end{itemize}

Table~\ref{fig:table sample} shows the properties and elemental abundances with respect to the Solar abundance values 
\begin{equation}
[X]=\log_{10}\bigg[\frac{X}{H}\bigg]_*-\log_{10}\bigg[\frac{X}{H}\bigg]_\odot \label{eq:4}
\end{equation}
with, $*$ representing the stellar value and $\odot$ the Solar value. The star denoted 0 is the Sun and serves as reference, and star 6 is a Red Giant\footnote{HIP 83359 is registered as a G5V star in the Hypatia Catalog which does not match its apparent surface temperature. It seems to be an error as \cite{Liu2010} catalogues it as a late Red Giant.}. It was decided to include this latter star in the sample in order to have at least one star displaying more extreme ratios of refractory elements.

The initial set of abundances used in the derivation of the condensation sequence is composed of 24 elements, but not all data is in the Hypatia Catalog. For some elements, we could fall back on established correlations like the one reported for sulphur and silicon \citep{Chen2002}; however, for many of the frequently missing elements (F, P, Cl, W), such correlations are less well established. Stellar elemental abundances could correlate based on their production sites \citep[e.g.\ different types of supernovae, merging neutron stars, see][for a recent review]{Matteucci2021}, but the subsequent dynamical evolution of the galactic interstellar medium may lead to significant spread in the abundance patterns \citep{Kobayashi2011a}. Our work aims at showing the diversity in mineralogy originating from typical elemental abundance spreads. Hence, to remedy for this purpose the issue of missing element abundances, we simply use the median of the depletion/enhancement of the known elements for that star. H and He are kept at the Solar abundance values for all the stars in the sample (see App.~\ref{sec:chemical}).

\subsection{Pressure-Temperature profile of the planet forming disk}
\label{sec:pt}

The pressure-temperature (P-T) profile that we use represents the Minimum Mass Solar Nebula \citep{Hayashi1981}. The temperature profile is assumed to be a power law, a good approximation taken from observations of planet forming disks \citep{williams2011}. We expect these profiles to evolve in time due to pre-main-sequence stellar evolution, variable accretion and changing dust opacities; however, we assume here a time-independent P-T profile and use it only to sample the respective physical conditions. We do not require for our further analysis that these conditions are associated with specific radial distances from the star. The derivation of the profile can be found in App.~\ref{sec:App proto disk}.

\begin{figure}
    \centering
    \includegraphics[width=9.5cm]{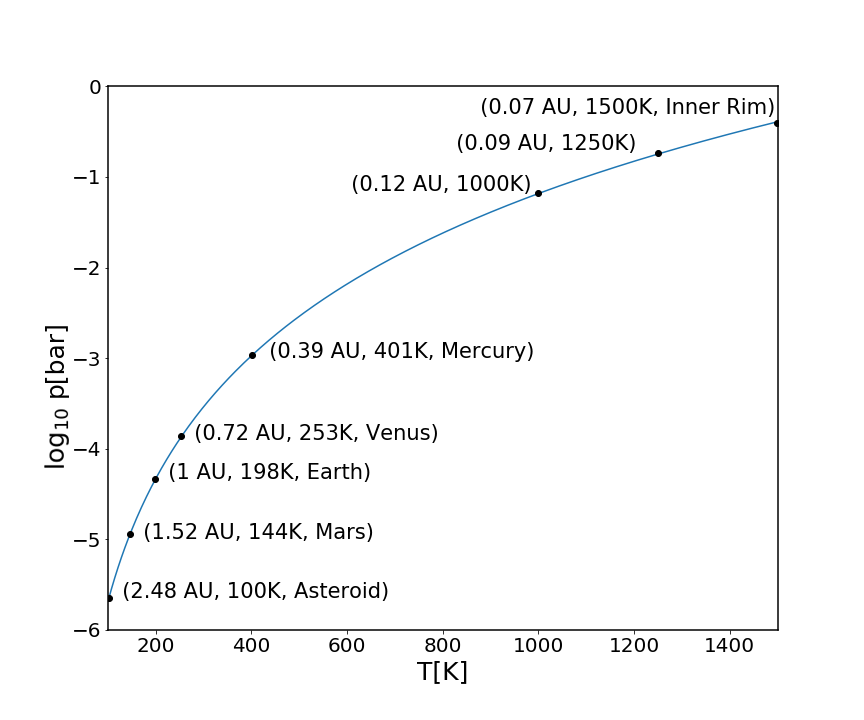}
    \caption{Pressure-Temperature profile in the mid-plane of the planet forming disk. Eight objects representing hypothetical rocky planets are shown as black filled circles and annotated with their respective distance from the host star and temperature. The host star would be located on the right.}
    \label{fig:pT profile}
\end{figure}

We sample the results of the condensation sequences by taking possible locations for the formation of hypothetical planets. Fig.~\ref{fig:pT profile} shows the positions of Mercury, Venus, Earth, Mars and a representative Asteroid. Their current radial positions might not be representative of the formation conditions of these bodies given the dynamical evolution of the proto-Solar nebula \citep[e.g.][]{raymond2020}. The terrestrial bodies in the solar system suggest that they formed at higher temperature than that used in this paper for their  respective location \citep{lewis1974}. 
Therefore, a further three hypothetical planets were chosen to sample the hotter end of the pressure-temperature profile. The highest one (1500 K) corresponds to the dust sublimation temperature (see App.~\ref{sec:App proto disk}) and is thus the point beyond which substantial mass can condense out and we can expect a significant solid reservoir for terrestrial planet formation.

\section{Results}
\label{sec:results}

In the following, we focus on a few interesting aspects of the condensation sequences and redirect the reader to App.~\ref{sec:condensate} for the full presentation of the results.

Two selection steps have been taken in order to investigate the influence of the main rock forming elements Mg, Fe, Si, as well as the more volatile O and S on the composition of forming planets. First, we look at a subselection of the condensation sequences (later only referred to as a subset) of the star sample for the element Mg and also for the elements Fe and S (the full set of condensates can be found in App.~\ref{sec:condensate}). The subsets for Si are not shown as the amount of condensates containing this element is too high to really be considered as a subset. However, if a specific condensate of Si is of interest, it will be included in the subsets of Mg.

Secondly, we look at specific positions in the P-T-plane. Solid bodies forming in those locations in situ will represent fictitious planets. As we have no reason to restrict ourselves to particular points on the P-T-plane, we investigate eight hypothetical terrestrial planets introduced in Section \ref{sec:pt}.

We first present the mineralogy of the Mg and the combined Fe and S subsets for the spread in elemental abundances from our 6 sample stars and then move on to discuss the implications for bulk planet composition assuming an in situ formation scenario. While we are aware that this is a strong assumption, the coupling of our results to more dynamical planet formation models will be part of subsequent work.

\subsection{Mineralogy}
\label{sec:mineralogy}


\subsubsection{Magnesium subsets}
\label{sec:Mg condensation}

\begin{figure*}
    \centering
    \resizebox{\hsize}{!}{
    \includegraphics{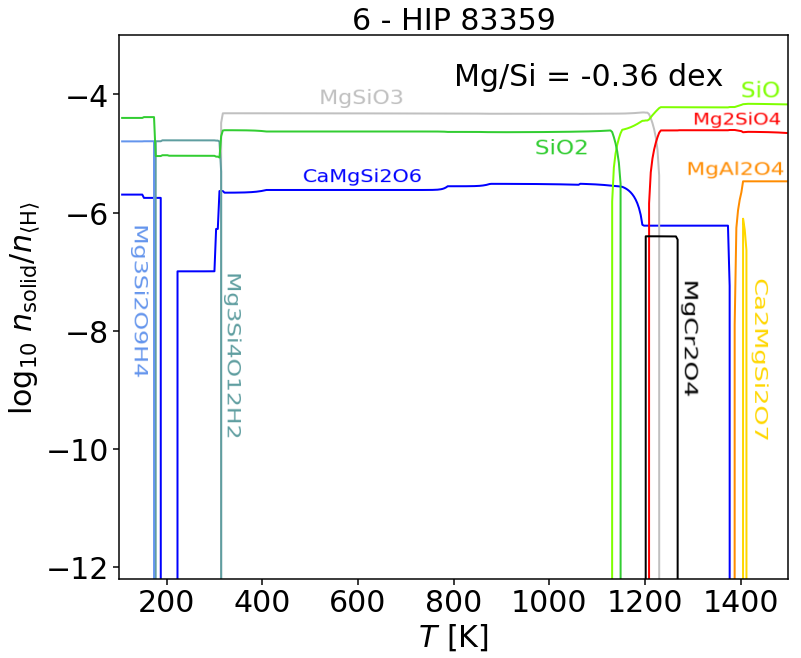}
    \includegraphics{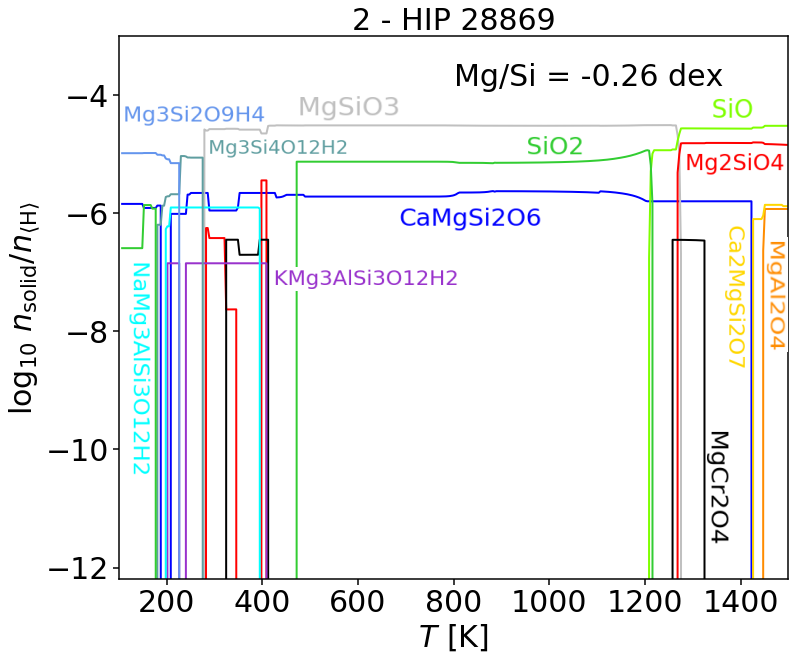}
    \includegraphics{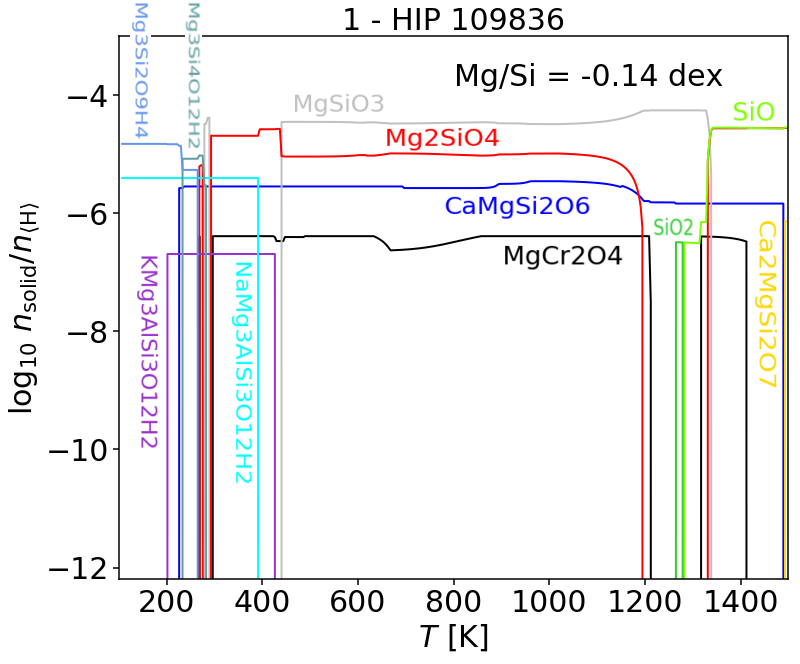}
    }
    \resizebox{\hsize}{!}{
    \includegraphics{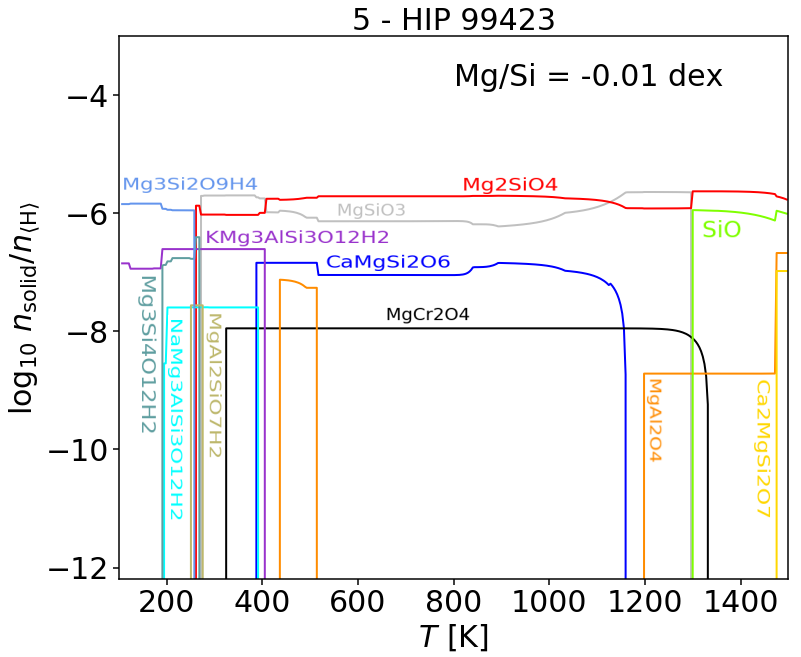}
    \includegraphics{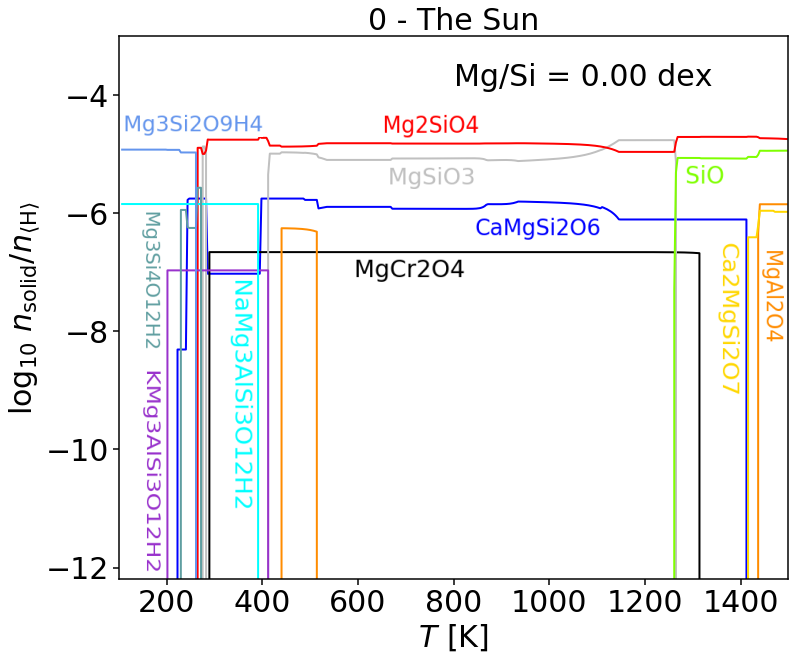}
    \includegraphics{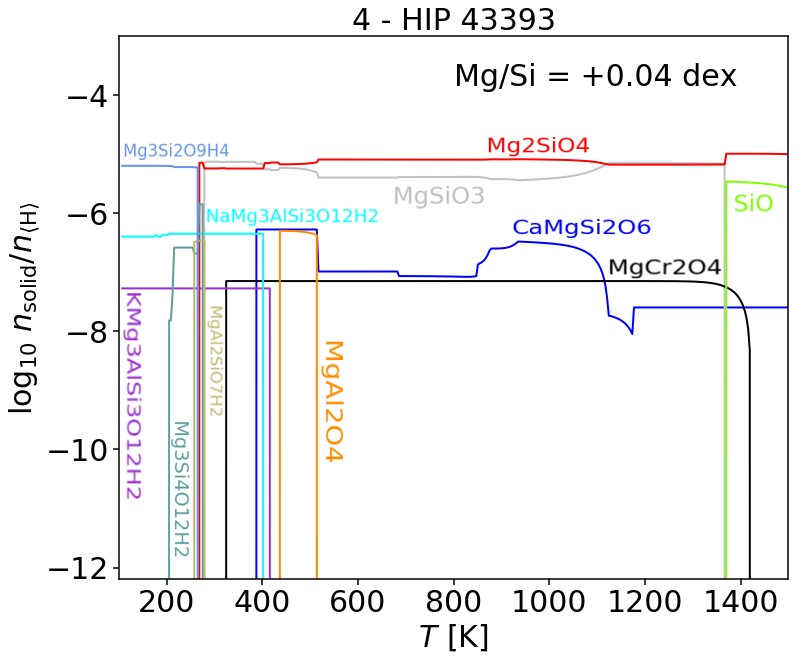}
    }
    \sidecaption
    \resizebox{\hsize/3}{!}{
    \includegraphics{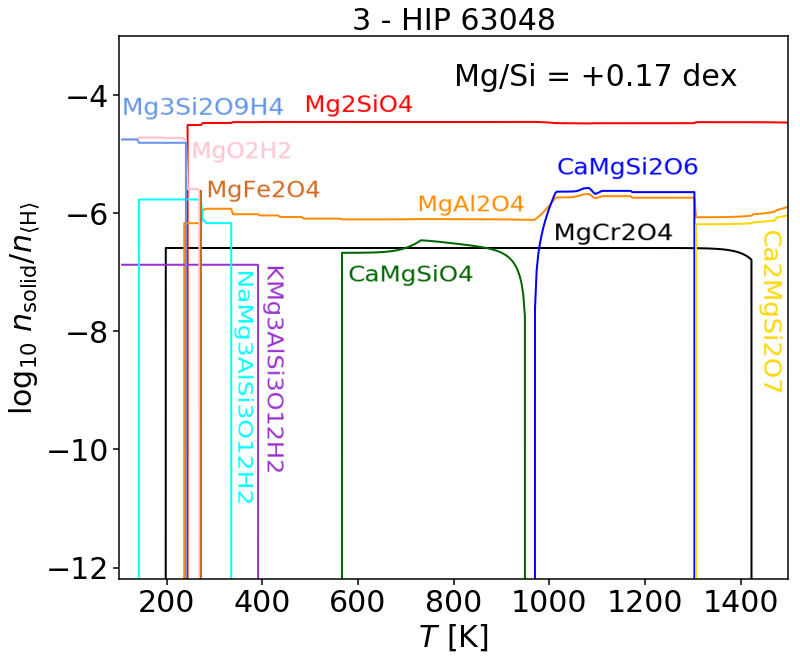}
    }
    \caption{Subsets of the condensation sequences of Mg for the 6 sample stars ordered in ascending value of Mg/Si at non constant pressure (see Fig.~\ref{fig:pT profile}). The ordering does not represent the total amount of solids. The full condensation sequences with all condensates can be found in App.~\ref{sec:condensate}. A shift is observed from low to high Mg/Si ratio where felsic rock-forming minerals disappear and mafic minerals emerge.}
    
    \label{fig:Mg1}
\end{figure*}

Fig.~\ref{fig:Mg1} shows the condensates for Magnesium. Some important condensates in planet composition are forsterite (Mg$_2$SiO$_4$), enstatite (MgSiO$_3$), SiO and Quartz (SiO$_2$) and these will be the focus of our analysis through the study of the Mg/Si ratio. So, if SiO and SiO$_2$ appear in the full condensation sequence, they will be represented in the subsequent Mg-subset. 

The Sun (Mg/Si$=\!0.0$~dex) is our reference star. For temperatures between $1250\!>\!T\!>\!1500$~K, Si is predominantly found in SiO (green curve in Fig.~\ref{fig:Mg1}) and forsterite  (red curve in Fig.~\ref{fig:Mg1}). Below 1250~K, there is an abrupt change where all SiO and part of the forsterite is transformed into enstatite (grey curve in Fig.~\ref{fig:Mg1}). These two condensates are then found coexisting in the temperature range $400\!<\!T\!<\!1250$~K. At temperatures between $\sim\!400$ and $\sim\!260$~K, enstatite disappears, leaving forsterite as the dominant Mg condensate. Below $\sim\!260$~K, more complex species form once the more simple forsterite starts to incorporate also hydrogen. This general behaviour is highly dependent on the value of the initial Mg/Si ratio. 

The two stars with a Mg/Si ratio close to solar ($+0.04$~dex and $-0.01$~dex for stars 4 and 5 respectively, see Fig.~\ref{fig:Mg1}) behave very similarly to the Sun. A main difference is that in these cases, forsterite and enstatite remain the main condensates down to $\sim\!~260$~K. 

A Mg/Si ratio of $-0.14$~dex (star 1) leads to an abrupt change in condensates around 1300~K. At this temperature, forsterite (red curve in Fig.~\ref{fig:Mg1}) and SiO (green curve in Fig.~\ref{fig:Mg1}) disappear and form instead enstatite (grey curve in Fig.~\ref{fig:Mg1}); Quartz (dark green curve in Fig.~\ref{fig:Mg1}) only appears in a very narrow temperature range ($1250\!<\!T\!<\!1280$~K). Below 1200~K, enstatite drops in abundance and we have a resurgence of forsterite.

The two stars with the lowest Mg/Si ratios of the sample ($-0.26$~dex and $-0.36$~dex, stars 2 and 6 respectively) extend the trends already seen in star 1. Forsterite (red curve in Fig.~\ref{fig:Mg1}) converts into enstatite (grey curve in Fig.~\ref{fig:Mg1}) and SiO (green curve in Fig.~\ref{fig:Mg1}) converts into SiO$_2$ (dark green curve in Fig.~\ref{fig:Mg1}). The main difference with star 1 being that Quartz now survives for an extended range of temperatures, down to 470~K for star 2. Below that it is converted into more complex condensates and then reemerges as Quartz at 180~K. For star 6, the lowest Mg/Si ratio, Quartz does not disappear, but instead becomes the main condensate (and Si carrier) for temperatures below 200~K.

The star with the highest ratio of Mg/Si ($+0.17$~dex, star 3) shows at the highest temperatures ($\sim\!1500$~K), forsterite (red curve in Fig.~\ref{fig:Mg1}) as the main condensate. This stays rather constant in abundance down to 240~K. At lower temperatures, more complex condensates form, again incorporating hydrogen. Interesting to note is the total absence of either enstatite , SiO or Quartz.

To summarise, when going from low to high Mg/Si ratios, we observe a shift with felsic rock-forming minerals disappearing and mafic minerals emerging.

\subsubsection{Iron and sulphur subsets}
\label{sec:Fe condensation}

\begin{figure*}
    \centering
    \resizebox{\hsize}{!}{
    \includegraphics{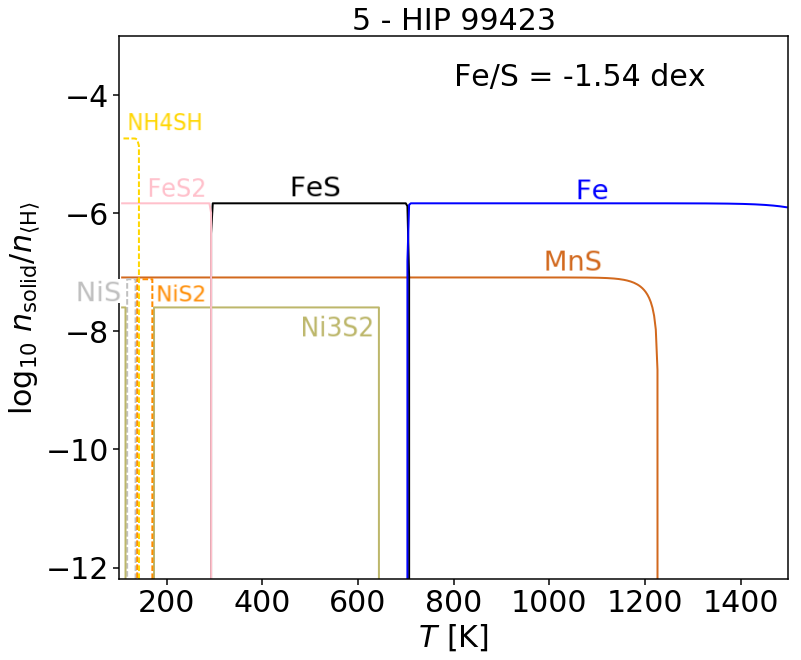}
    \includegraphics{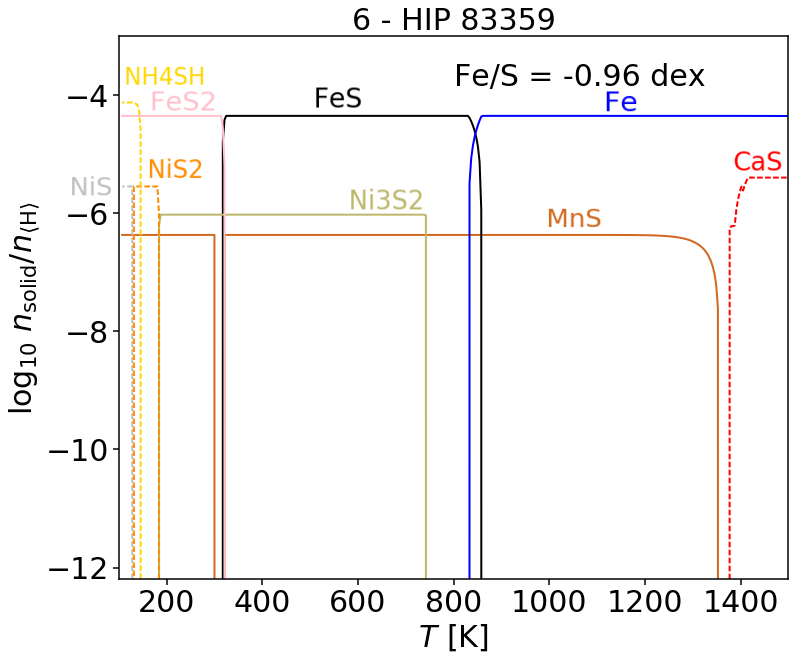}
    \includegraphics{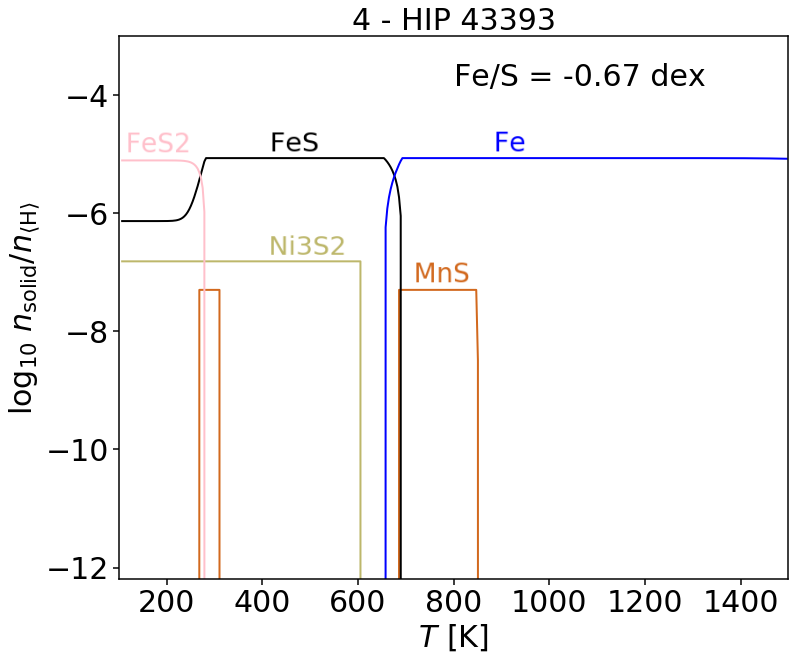}
    }
    \resizebox{\hsize}{!}{
    \includegraphics{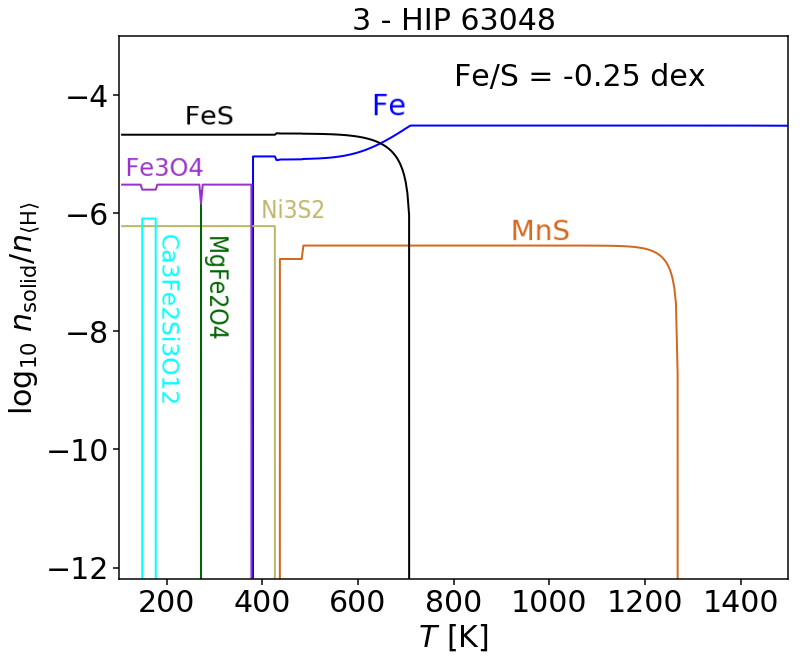}
    \includegraphics{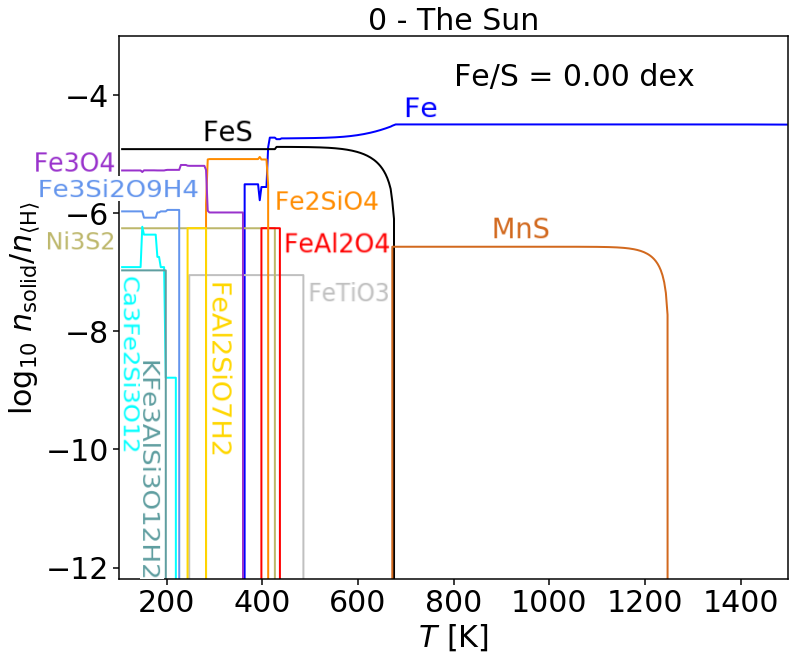}
    \includegraphics{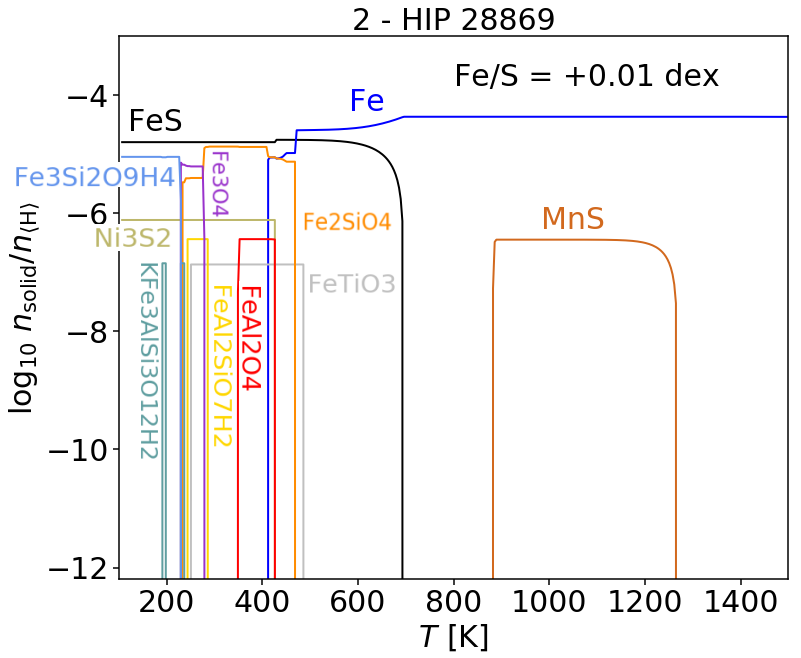}
    }
    \sidecaption
    \resizebox{\hsize/3}{!}{
    \includegraphics{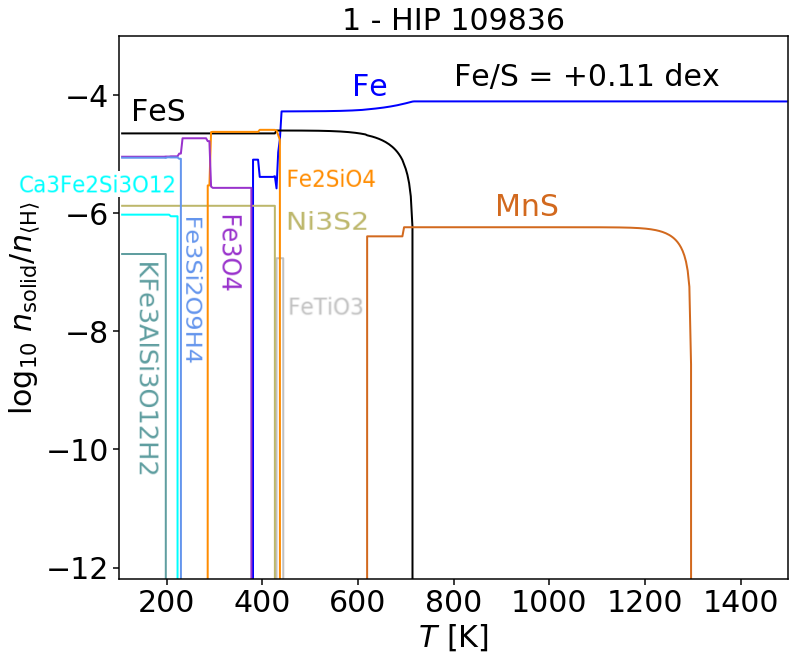}
    }
    \caption{Subsets of the condensation sequences of Fe for the 6 sample stars ordered in ascending value of Fe/S at non constant pressure (see Fig.~\ref{fig:pT profile}). The ordering does not represent the total amount of solids. The full condensation sequences with all condensates can be found in App.~\ref{sec:condensate}. Fe denotes here metallic iron.}
    \label{fig:Fe1}
\end{figure*}

The condensates containing iron (Fig.~\ref{fig:Fe1}) are striking in their simplicity (relative amount of condensate species) when compared to those of magnesium. Iron is one of the few elements that is fully condensed out at the inner rim of the planet forming disk (1500~K) for all stars in the sample i.e.\ independent of the element abundances. 

The first, more complex, condensate that forms from metallic iron is FeS around 700 K, slightly depending on the star (elemental abundances). One exception is star 6 (Fe/S$=\!-0.96$~dex), where the transition occurs at 858~K. The type of the new condensates forming at temperatures below that is then highly dependent on the Fe/S ratio and we discuss this in the next paragraphs going from high to low Fe/S ratio. 

A high Fe/S ratio implies that there are relatively fewer S atoms readily available to bind with Fe. The fewer S is available with respect to Fe, the lower the temperature at which Fe gets incorporated into FeS. For stars 0 to 3 (Fe/S$\geq\!-0.25$~dex), the depletion of iron is a relatively gradual process over a few 100~K which leads to the total disappearance of metallic Fe only when Fe-bearing silicates (Fe$_2$SiO$_4$) or magnesium ferrite (MgFe$_2$O$_4$) form (below $\sim\!400$~K). Sulphur only forms two other types of condensates, namely MnS and Ni$_3$S$_2$.

Stars 4 to 6, which are associated with the lowest Fe/S ratios of the star sample, show the most extreme behaviours. The amount of S readily available is such that virtually all the Fe (blue curve in Fig.~\ref{fig:Fe1}) condenses at high temperatures ($\gtrsim\!700$~K) and stays stable as FeS (black curve in Fig.~\ref{fig:Fe1}) down to $\sim\!300$~K ($321-279$~K depending on the Fe/S ratio). Below these temperatures FeS combines further with S leading to FeS$_2$ (pink curve in Fig.~\ref{fig:Fe1}). As a matter of fact, iron is entirely consumed into FeS and FeS$_2$ and does not get incorporated into silicates. The leftover sulphur goes on forming more MnS around 300~K for star 4, and new condensate species for the stellar systems 5 and 6. Star 6 is unique in this sample as sulphur is already present as CaS at the inner rim of the planet forming disk.

This leads to the division of our stellar sample into two groups. One group with a Fe/S ratio $\geq\!-0.25$~dex that forms, at low temperatures, Fe-bearing silicates and another group with a Fe/S ratio $\leq\!-0.67$~dex forming only FeS, FeS$_2$ ,and at low temperatures more complex sulphur condensates.

\subsection{Planet composition}
\label{sec:planet composition}

In order to derive the bulk chemical composition of a planet we need to take another step after the derivation of the condensation sequences. We need to find the abundance of each of the 24 chemical elements selected in our experiment within the condensates. The key goal of this paper is to look at the differences in the chemical composition of a hypothetical solid body around a star in which the abundances of refractory elements vary with respect to the solar values. 

After the condensation of solids within the planet forming disk, the various condensates still need to go on forming a planet either through core accretion \citep{Pollack1996} or gravitational instabilities \citep{Boss1997}. We note that gravitational instabilities occur in the outer regions of massive disks, resulting in massive gas giant planets. This planet formation scenario is therefore less relevant for our study. During the planet formation process, pressures and temperatures become high enough to sublimate the condensates within the forming planetesimals/protoplanets \citep[e.g.][]{Fu2014}. Thus, what matters is not the specific identity of the condensates (e.g.\ MgO, FeS, H$_2$O) that get incorporated into this body but rather the abundance of each of the elements (H, He, Li, ...) as well as the ratios of these elements. We want to assess to which extent the original gas-phase element ratio is conserved in the condensed phase as a function of temperature. 

We have a specific abundance $\epsilon_j$ for a given condensate $j$ and stoichiometry for the elements that form that condensate. For a condensate with 3 elements A, B, C, and respective stoichiometry $a_j$, $b_j$, $c_j$, this becomes A$_{a_j}$B$_{b_j}$C$_{c_j}$. The abundance of the constituent/element A within the solid phase is then given by the sum over all condensates containing A
\begin{equation}
\epsilon({\rm A}) = \sum_j a_j \epsilon_j    
\end{equation}
In the following subsections, we show the impact of changing the ratios Mg/Si, Fe/Mg and Fe/S (App.~\ref{sec:chemical}) on the hypothetical planets that we form in situ. For this, we use the mass fraction (wt$\%$) of the forming planets and the element ratios from the gas phase to the condensed phase. Another quantity we derive are the ratios of oxygen with respect to the refractory elements to help us understand the boundary between an oxidized or reduced interior and Fe/S to help us understand the S wt$\%$ present in the core of a hypothetical planet.

\subsubsection{Linking stellar and planet elemental composition} 
\label{sec:elementratios_in_planets}

\begin{figure*}
    \centering
    \resizebox{\hsize}{!}{
    \includegraphics{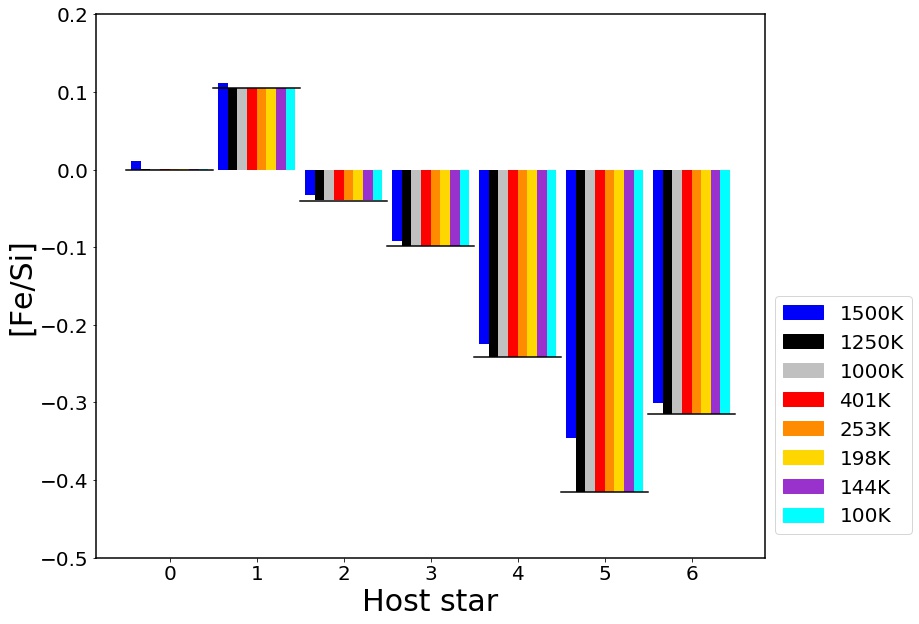}
    \includegraphics{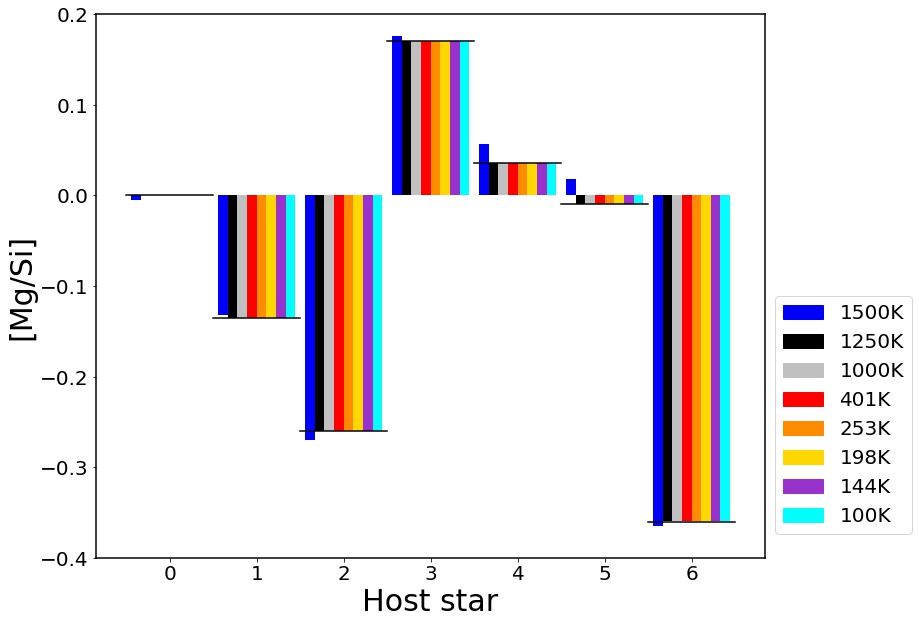}
    }
    \resizebox{\hsize}{!}{
    \includegraphics{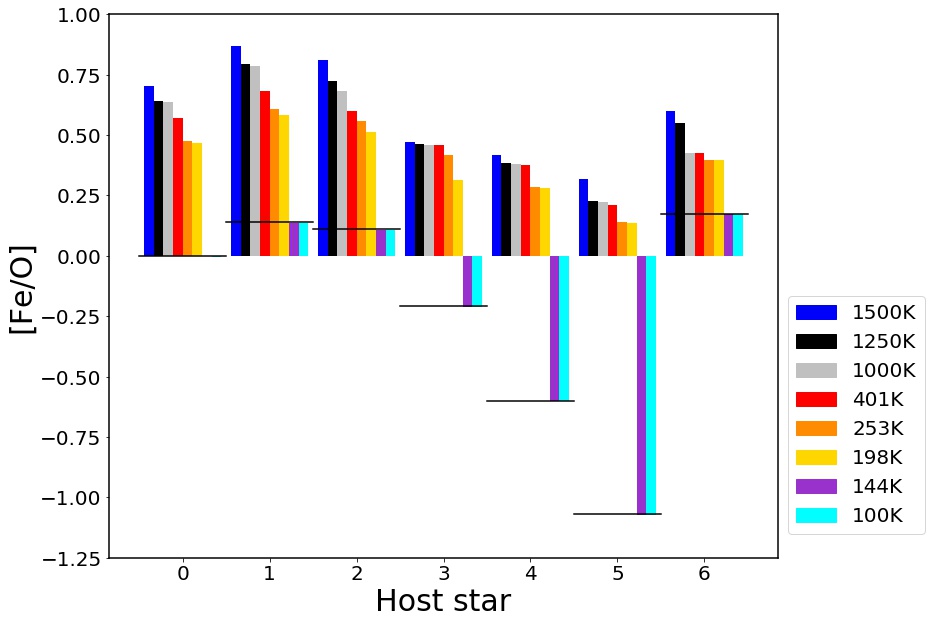}
    \includegraphics{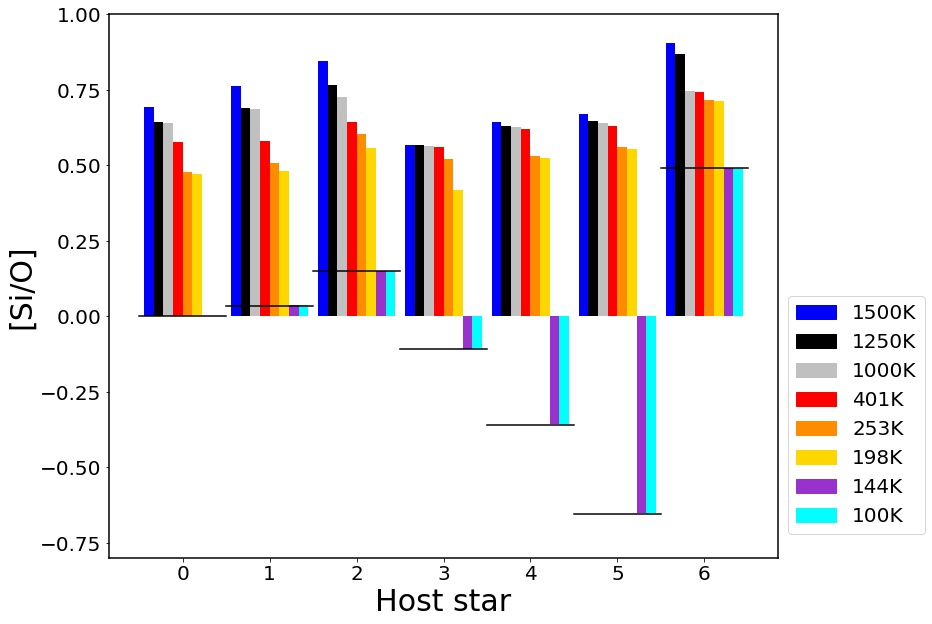}
    }
    \resizebox{\hsize}{!}{
    \includegraphics{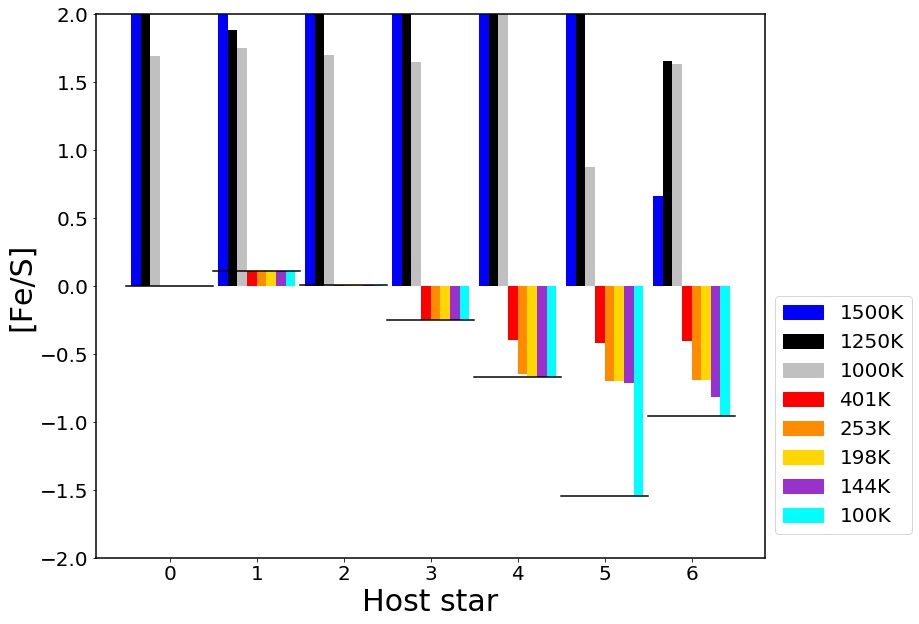}
    \includegraphics{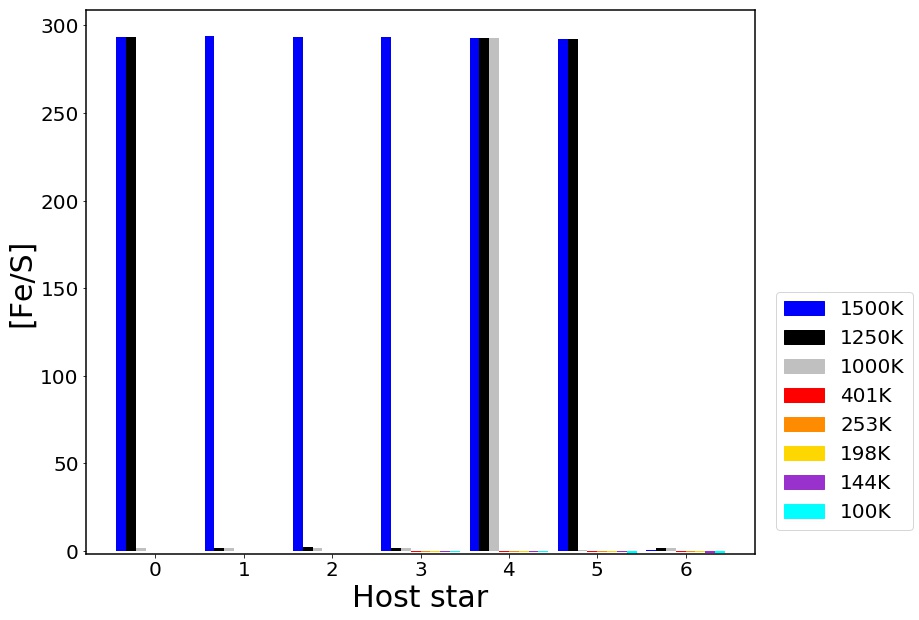}
    }
    \caption{The ratios found in the condensed phase from which the hypothetical planets form. The ratios are normalised by the Solar abundance values. The x-axis labelled from 0 to 6 represents the host star where the planets formed. A black horizontal line is drawn to represent the initial value of the ratio in the gas phase. Bottom right panel is an extension of the bottom left panel.}
    \label{fig:chart ratio}
\end{figure*}

The refractory elements Mg, Si and Fe completely condense, as can be seen from Fig.~\ref{fig:chart ratio}. Only the planet at 1500~K shows a small departure from the gas phase ratio (|Fe/Si|\,$\leq \!0.06$~dex and |Mg/Si|\,$\leq\!0.03$~dex). The incomplete condensation of the more volatile elements is a well established fact \citep[e.g.][]{Woitke2018}. In this simple in situ scenario, planets forming inside the snowline contain less than the original gas-phase oxygen abundance. Sulphur condenses in our models as FeS and FeS$_2$ at temperatures that depend on the initial gas-phase elemental abundance ratios; therefore, the difference between the initial gas-phase Fe/S ratio and that in the condensates varies within the hypothetical planets around each star of our sample.

When moving now from element ratios in the condensed phase (within the planet) to mass fractions (Fig. \ref{fig:chart}), we see that the snowline plays a key role. Planets forming outside the snowline contain $>50$~wt\% of water ice (so oxygen), the only exception being star 6. The incorporation of volatiles starts already around 400~K where phyllosilicates appear in the condensation sequence (Fig.~\ref{fig:Mg1}). Phyllosilicates can also play a role in 'delivering' water to terrestrial planets, especially the interiors. This has been studied for example by \citet{DAngelo2019} and \citet{Thi2020}; these phyllosilicates are thermally more stable than water ice and hence not subject to the same degree of devolatilization in the planets envelope as icy pebbles \citep{Johansen2021}. The formation of such phyllosilicates and their incorporation into planets leads to a relatively lower mass fraction of refractory elements such as Mg, Si, and Fe in planets forming below $\sim\!400$~K. We also need to keep in mind that the absolute abundances become important in setting the absolute wt\%. The next subsections discuss the key results concerning the mass fractions of our hypothetical planets.

\subsubsection{Magnesium and Silicon mass fractions}

\begin{figure*}
    \centering
    \resizebox{\hsize}{!}{
    \includegraphics{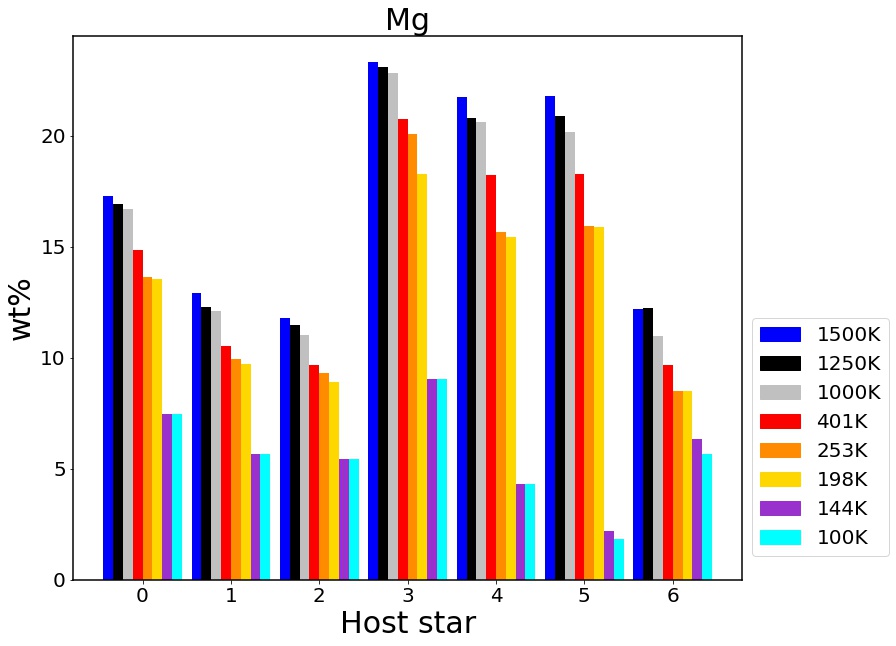}
    \includegraphics{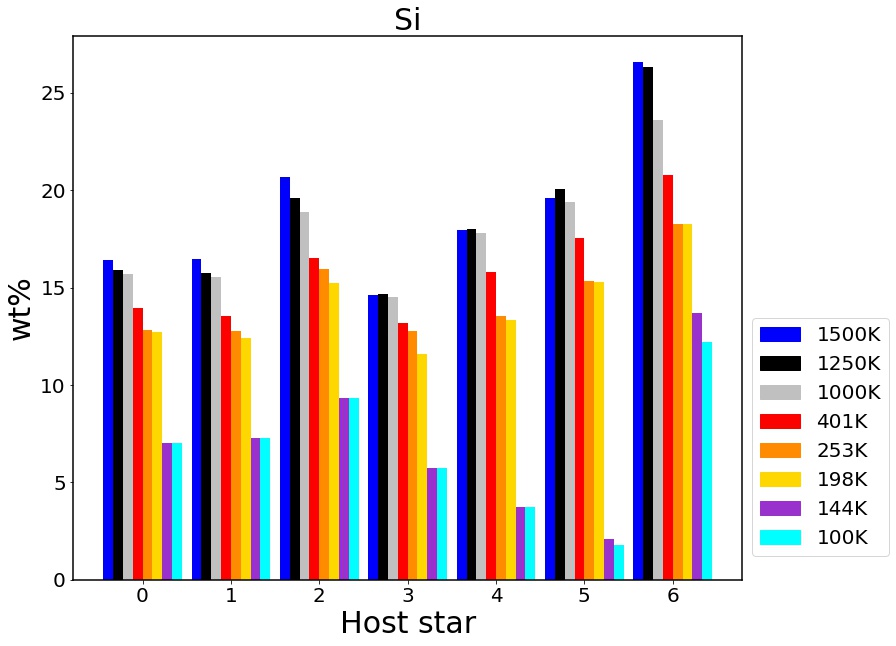}
    }
    \resizebox{\hsize}{!}{
    \includegraphics{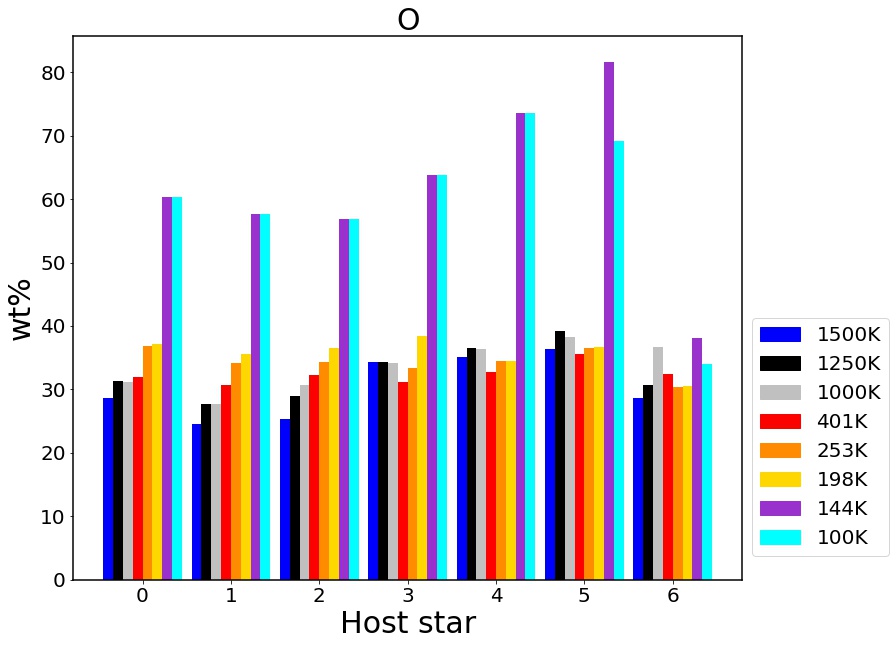}
    \includegraphics{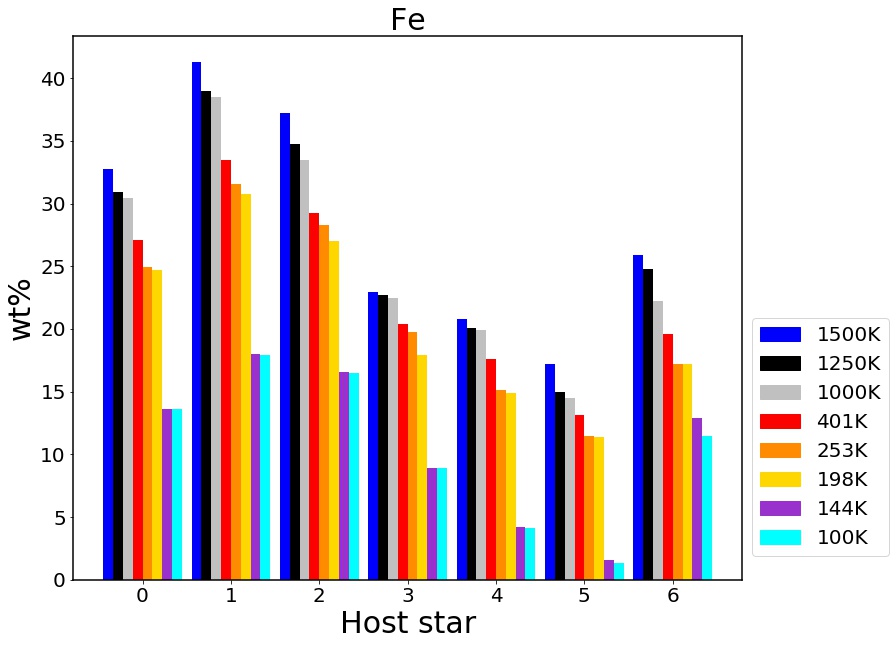}
    }
    \sidecaption
    \resizebox{\hsize/2}{!}{
    \includegraphics{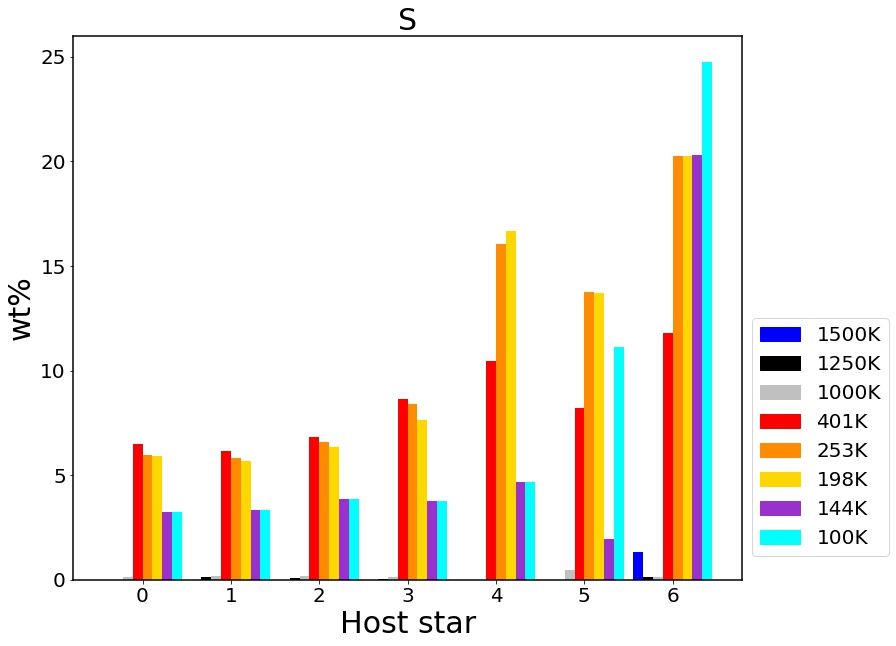}
    }
    \caption{The mass fraction of the hypothetical planets forming in situ. Values are given as percentage of the total mass of the body. The missing mass fraction necessary to add up to 100$\%$ is contained in the chemical elements not discussed here. The x-axis labelled from 0 to 6 indicates the host star where the planets formed.}
    \label{fig:chart}
\end{figure*}

From the condensation sequences (Sect.~\ref{sec:mineralogy}), we find that Mg and Si are intricately linked and, in association with other metals, form various condensate species. Moreover, we saw in Sect.~\ref{sec:Mg condensation}, that the Mg/Si ratio has an influence on the formation of silicate and quartz condensates. The link between the different initial abundances of the host stars and the mass fractions of Mg and Si is not trivial. However, it is clear from Fig.~\ref{fig:chart} that a deviation from solar abundances has a large impact ($>\!10$~\% in some cases) on the mass fractions. 

It is clear that within a stellar system, the lower the temperature, the lower the Mg-, or Si-mass fraction. This is linked to the formation of phyllosilicates around 400~K (planets forming at 401~K, 253~K, and 198~K) and the crossing of the snowline for planets forming at lower temperatures (144~K and 100~K). Past these limits, the presence of O becomes prevalent in the condensate mixture, leaving the refractory elements such as Mg and Si with lower mass fractions.

\subsubsection{Oxygen mass fraction}

The element ratios involving oxygen (Fig.~\ref{fig:chart ratio}) show that planets forming well inside the snow line (at temperatures equal or higher than 401~K) do so in a disk region where a high fraction of the oxygen is in the gas phase. Hence, they end up forming from a condensed phase with higher Fe/O and Si/O ratios, while outside the snow line (at 144~K and 100~K) planets inherit ratios equal to that of the initial gas phase due to the efficient formation of water ice \citep[e.g.][]{Woitke2018}. Planets forming at intermediate temperatures of 198~K  and 253~K (so inside the snowline) partially inherit the gas phase O abundance due to the formation of phyllosilicates for temperatures below $\sim\!400$~K.

For the stars 0 through 2, we have a gradual increase in the oxygen mass fraction with temperature and a steep jump at the snowline (Fig.~\ref{fig:chart}). Stars 3, 4 and 5, on the other hand, show an almost constant oxygen mass fraction for planets forming inside the snowline ($35\,\pm\!5\,\%$). This behaviour is also seen for the two planets outside the snowline for star 6. The oxygen mass fraction is affected here by other species present in the condensed phase. The planets forming in the stellar system 6 can be explained by the high initial abundances of refractory elements as well as sulphur present in the gas mixture ([Fe]$=\!+0.145$, [Mg]$=\!+0.1$, [Si]$=\!+0.46$, and [S]$=\!+1.1$). 

A one-to-one correlation between the original stellar abundances and the oxygen mass fraction is not apparent. While oxygen varies from $-0.265$~dex (star~5) to $+0.25$~dex (star~1), the mass fraction does not reflect this change. Indeed, star~5 with the lowest abundance of oxygen with respect to the Sun, has one of the highest oxygen mass fraction. This is due to the even lower abundance (wrt solar) of all others elements within the stellar system (see Table~\ref{fig:full chem}). This makes very clear that the final composition (mass fractions) of a planet is strongly affected if the full observed abundance pattern of the stellar system is taken into consideration.

\subsubsection{Iron and Sulphur mass fraction}


Like earlier work, we also find that the iron mass fraction inside a planet is directly correlated with the iron abundance in the initial gas-phase (stellar abundance Table~\ref{fig:full chem}). The main new aspect of our work is the important role of sulphur in binding iron. We saw in Fig.~\ref{fig:Fe1} that the silicates forming around stars~4, 5 and 6, which have a very low Fe/S ratio,  did not incorporate any iron. This is due to sulphur taking up all iron and forming FeS and FeS$_2$. As a consequence, planets forming around these stars show a very high wt\% of sulphur (Fig.~\ref{fig:chart}), in some extreme cases up to 20-25\% (star~6). We will come back to the implications of this in Sect.~\ref{sec:discuss-planet-composition}.

For stars with low Fe/S ratio, we also see that many of the hypothetical planets --- especially the warmer ones closer to the star --- would not inherit the original gas phase Fe/S ratio (Fig.~\ref{fig:chart ratio}). To elaborate on this, in star~4, planets forming at temperatures $T\!\leq\!198$~K have the same Fe/S ratio as the one found in the initial gas phase, while in star~5 and 6 only the planet at $T\!=\!100$~K shows this behaviour. It is directly linked to the ability of sulphur to efficiently form at low temperatures other sulphuric condensates. A specific example for this is the planet forming at $T\!=\!144$~K in star~6: At $T\!<\!150$~K, NH$_4$SH condenses out and becomes the prevailing sulphur bearing condensate. Therefore, only planets forming at these temperatures can inherit the full gas phase element ratio. The lower the stellar Fe/S ratio is, the more complex sulphur bearing condensates form at low temperatures (see again Fig.~\ref{fig:Fe1}). This shifts the sulphur condensation line to lower temperatures.

\section{Discussion}
\label{sec:discussion}

We compare here our results with previous work, discuss the limitations of our approach and the implications our results have on planet composition.

\subsection{Comparison with previous work}
\label{sec:comparison}

\begin{figure}
    \hspace*{-2mm}
    \includegraphics[scale=0.38]{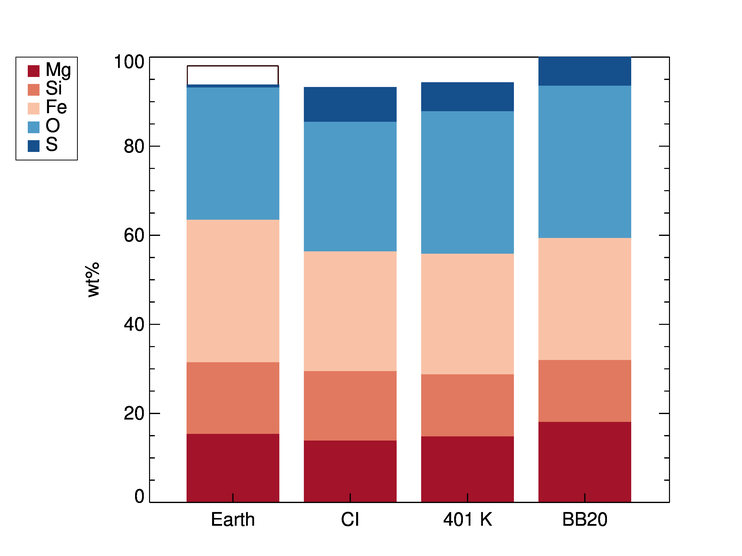}
    \vspace*{-6mm}
    \caption{Comparison of the atomic wt\% of the key elements in the bulk Earth and CI chondrites \citep{mcdonough2021}, our 401~K planet around the Sun, and the work of \citet[][BB20]{Bitsch2020}. The open box for bulk Earth indicates the uncertainty in the sulphur, which could be up to 4.2~wt\% (Sect.~\ref{sec:discuss-planet-composition}).}
    \label{fig:compare}
\end{figure}

It is interesting to compare our results to bodies in the Solar System, such as bulk Earth and CI chondrites \citep{mcdonough2021}. Figure~\ref{fig:compare} shows that the composition we find from our equilibrium condensation for a planet forming at 401~K around the Sun compares very well to that of bulk Earth.


It is also interesting to compare our work to that of previous studies. The chemical model of \citet{Bitsch2020} is more simple than the one we used, and they only consider solid planetary building blocks forming on each side of the snowline for \mbox{sub-,} super-, and solar metallicities (scaling the logarithm of all element abundances using a linear relationship with the logarithm of the Fe abundance). This makes it somewhat complicated to compare all cases we have developed here. However, the solar case can be compared to our work. The mass fractions resulting from their study for Fe, Si, O and S are very similar to our planet forming at 401~K (Fig.~\ref{fig:compare}). We find a lower Mg wt\%, which actually brings our results closer to the value of the bulk Earth found in \citet{McDonough1995} when considering 15\% of light elements in the core. This also means that to have an Earth-like composition from the outcome of equilibrium chemistry, the planet would have had to form at a higher temperature closer to $T\!=\!401$ K, rather than the temperature we use at 1~au ($T\!=\!198$~K) in our disk model. This can be easily explained when taking into account that the solar luminosity had been higher in the past and thus the disk was likely hotter \citep[e.g.][]{DAngelo2019}. Also high accretion rates early on will produce a much warmer disk due to viscous heating \citep[e.g.][]{Dullemond2007,Min2011}.

We have used an in situ approach to calculate bulk planet compositions. Both migration and gravitational scattering of planetary embryos will modify the final planet composition. More detailed studies are needed to quantify these effects in the context of our chemical model. We expect that the snowline will be an important parameter affecting the final outcome, as is the case for our in situ models. For instance, the inward migration of icy planetary cores may result in more water-rich close-in planets than our in situ models predict \citep{Raymond2018,Bitsch2020}. In general, such dynamical formation scenarios increase the diversity of the final composition of inner disk planets \citep{Raymond2018}.  



\subsection{Limitations}
\label{sec:limitations}

The condensation sequences are an interesting tool to analyse what type of condensates form on the basis of the initial abundance of each chemical element. We have seen that changes in the Mg/Si ratio result in differences in forsterite, enstatite and Quartz abundances, or that the Fe/S ratio dictates the possible condensates that Fe can form. However, it is important to remember that the condensation sequences are generated by the GGchem code for a specific pressure-temperature profile. It calculates the abundance of the condensates to expect when an initial cloud cools down. For the purpose of planet building, an extra step is needed. These condensates need to form the final planet through a process of growth; for this study, core accretion following \citet{Pollack1996} is the most relevant. In that process, temperatures and pressures increase such that the identity of the condensate species is lost. What remains is the individual chemical elements. To draw more conclusions about the final planet structure on the basis of the initial refractory ratios found in the cloud, one would need to model the accretion and subsequent cooling of the planet. This would also involve radial transport processes of solids in such a planet forming disk, so a non-local component \citep[e.g.][]{Ciesla2011,raymond2020,Lichtenberg2021}.

However, the condensation sequences are still a good approximation as to what to expect as the ratios of the refractory elements Fe/Si and Mg/Si (Fig.~\ref{fig:chart}) are the same as in the gas phase. Clearly, this is not the case for more volatile elements such as oxygen and sulphur. Here, non-equilibrium processes during the cooling will play a role \citep[see e.g.][]{Herbort2021}.



Finally, it is also important to ask how reliable the initial chemical abundance data are, which are used to run the GGchem code. These were taken from the Hypatia Catalog, and as mentioned in Sect.~\ref{sec:abund}, the data collected in the Hypatia Database are from various sources, which employed very different methods to derive them. The final values presented for each star are actually the median of all recorded sources, which in some cases, show discrepancies from one another larger than the error bar of each individual value. \citet{Unterborn2016} show that the derivation of stellar abundances is not a trivial exercise and can lead to discrepancies that cannot only be explained by the difference in approach used by the different teams. It is important to keep that in mind when putting our results into a larger context.

\subsection{Planet interior structure}
\label{sec:discuss-planet-composition}

Both the mass fraction and the ratios of the various chemical elements in the condensed phase of the planet forming disk are important in understanding the interior structure of the planets that are forming from those building blocks. The composition of the core will influence its size \citep[e.g.][]{valencia2007}, the magnetic field of the planet \citep[e.g.][]{Noack2020} or change the heat flow from the core to the surface \citep{Schubert2001}. Clearly, the mantle mineralogy depends on the various refractory element ratios \citep[see also][]{Putirka2019} and this can have important consequences for the long-term thermal evolution and volatile cycling (e.g.\ H$_2$O and CO$_2$) of the planet \citep{spaargaren2020}.

Iron is a very important element as its heavy weight will lead it to fractionate and form the core of the forming planet. However, the core is not only composed of pure Fe. 
Seismic studies \citep[e.g.][]{Birch1952} show that the Earths core has a lower density than expected based on pure Fe-Ni alloy, indicating the presence of lighter elements. In recent years sulphur became a prime light element candidate for explaining the low density in the core \citep{Hirose2013}. Therefore, the behaviour of Fe and S is important to describe the percentage of lighter alloys present in a planets core. Table~1 of \citet{Hirose2013} reports the mass fraction of Si, O, S, C and H that might be present in the Earth core from various sources. The possible S wt$\%$ in the core ranges from 0 to 13\% depending on the source; this translates into an uncertainty of 4.2~wt\% for bulk Earth (see Fig.~\ref{fig:compare}). A large fraction of the core can then be composed of sulphur. Given our finding that the Fe/S influences, in some cases drastically, the bulk planet S wt\% we can speculate about the interior structure resulting for a planet assembling under such conditions. Star~6 is the most striking example. Its Fe/S ratio leads it to only form from metallic Fe the condensates FeS and FeS$_2$ leading for planets forming at $T\!\leq\!253$~K to a bulk planet S wt\% larger than the Fe wt\%. Depending on how much sulphur actually sinks into the core, it could lead to sulphur being the predominant element in the planet core. 

The oxidation state of the material that is accreted to a growing planet sets the boundary conditions for the formation of an Fe core. We have shown that Fe can be accreted in reduced form, i.e. metallic Fe or FeS, or in an oxidised form, and that the occurrence of these materials depends on location in the disk as well as on the parent star chemical composition. Planets formed in the inner, dry disk have a more reducing chemical composition, while planets formed beyond the snowline have a larger intake of oxygen. Oxidised forms of Fe are considered to be lithophile and remain in the mantle, while Fe and the calcophile FeS can accrete to the core.  Upon accretion, the reduced forms of Fe can become oxidised when mixed into the growing parent body. This can happen if the timescale for oxidation is shorter than the timescale for Fe to accrete to the core, possibly leading to coreless planets (Elkins-Tanton et al. 2008). As mentioned above, in some of our models no oxidised Fe is present in the solids, which may lead to a large Fe core, possibly containing a high fraction of sulphur. For Mars, where the core pressure is much lower than for Earth, \citet{Urakawa2004} suggest that the possibility to form a stable core depends on the wt\% of sulphur in the core. 

The key result of this work is that a change in the initial chemical abundance of the gas phase, even a small one (less than a factor two), significantly changes the outcome of the condensation sequence. This leads to planet compositions with large differences in their Mg, Si, Fe, O and S mass fraction. The strong implications of this warrant further and more detailed analysis of the planet formation phase. 

\section{Conclusions}
\label{sec:conclusions}

We have used the GGchem code to investigate how the equilibrium condensation sequence changes when the ratios of the refractory elements Mg, Si and Fe are changed from the Solar abundance values according to the spread of elemental abundances found in the Hypatia Catalog. We have studied this for typical pressure-temperature conditions found in a planet forming disk and investigated the composition of hypothetical terrestrial planets forming around the sample of stars. 

We find that changing the Mg/Si ratio has a direct impact on the type of silicates formed from the condensation, with a shift from forsterite (Mg$_2$SiO$_4$) and SiO to enstatite (MgSiO$_3$) and Quartz (SiO$_2$) when the Mg/Si ratio decreases. Furthermore, changes in the Fe/S ratio lead to two types of condensation sequences. For Fe/S$\geq\!-0.25$~dex, Fe forms various types of Fe-bearing silicates at low temperatures while for Fe/S$\leq\!-0.67$~dex, Fe only forms FeS and FeS$_2$.

Ratios of refractory elements are found to directly translate from the gas phase to the condensed phase. However, elemental ratios with respect to O show that planets inside the snowline form from a condensed phase with larger ratios Fe/O and Si/O than in the gas phase. Oxygen is only fully condensed outside the snowline, and as such, any planet forming inside the snowline will have a higher Fe/O ratio. This will lead to planets having more reduced interior. In addition, the Fe/S ratio directly impacts the amount of S wt$\%$ in the planet.

In conclusion, our study shows that the typical spread ($\sim\!0.1-0.2$~dex) in the abundance of refractory elements found in main sequence G-type stars in the Solar neighbourhood can impact significantly the outcome of planet formation. Planet formation scenarios should include the chemical abundance data of the host star and not impose solar abundance values to study bulk exoplanet properties.

\begin{acknowledgements}
I.K.\ and P.W.\ acknowledge funding from the European Union H2020-MSCA-ITN-2019 under Grant Agreement no.\ 860470 (CHAMELEON). This research was supported and inspired by discussions through the ISSI (International Space Science Institute) International Team collaboration 'Zooming In On Rocky Planet Formation' (team 482).
\end{acknowledgements}

\bibliographystyle{aa} 
\bibliography{aa.bib} 

\begin{thebibliography}{57}
\expandafter\ifx\csname natexlab\endcsname\relax\def\natexlab#1{#1}\fi

\bibitem[{Adibekyan {et~al.}(2021)Adibekyan, Dorn, Sousa, Santos, Bitsch,
  Israelian, Mordasini, Barros, Mena, Demangeon, Faria, Figueira, Hakobyan,
  Oshagh, Kunitomo, Takeda, Jofré, Petrucci, \&
  Martioli}]{adibekyan2021chemical}
Adibekyan, V., Dorn, C., Sousa, S.~G., {et~al.} 2021, The Chemical link between
  stars and their rocky planets

\bibitem[{{Adibekyan} {et~al.}(2015){Adibekyan}, {Santos}, {Figueira}, {Dorn},
  {Sousa}, {Delgado-Mena}, {Israelian}, {Hakobyan}, \&
  {Mordasini}}]{Adibekyan2015}
{Adibekyan}, V., {Santos}, N.~C., {Figueira}, P., {et~al.} 2015, \aap, 581, L2

\bibitem[{{Aguichine} {et~al.}(2020){Aguichine}, {Mousis}, {Devouard}, \&
  {Ronnet}}]{Aguichine2020}
{Aguichine}, A., {Mousis}, O., {Devouard}, B., \& {Ronnet}, T. 2020, \apj, 901,
  97

\bibitem[{{Armitage}(2019)}]{Armitage2019}
{Armitage}, P.~J. 2019, Saas-Fee Advanced Course, 45, 1

\bibitem[{{Asplund} {et~al.}(2009){Asplund}, {Grevesse}, {Sauval}, \&
  {Scott}}]{Asplund2009}
{Asplund}, M., {Grevesse}, N., {Sauval}, A.~J., \& {Scott}, P. 2009, \araa, 47,
  481

\bibitem[{{Audard} {et~al.}(2014){Audard}, {{\'A}brah{\'a}m}, {Dunham},
  {Green}, {Grosso}, {Hamaguchi}, {Kastner}, {K{\'o}sp{\'a}l}, {Lodato},
  {Romanova}, {Skinner}, {Vorobyov}, \& {Zhu}}]{audard2014}
{Audard}, M., {{\'A}brah{\'a}m}, P., {Dunham}, M.~M., {et~al.} 2014, in
  Protostars and Planets VI, ed. H.~{Beuther}, R.~S. {Klessen}, C.~P.
  {Dullemond}, \& T.~{Henning}, 387

\bibitem[{{Birch}(1952)}]{Birch1952}
{Birch}, F. 1952, \jgr, 57, 227

\bibitem[{{Bitsch} \& {Battistini}(2020)}]{Bitsch2020}
{Bitsch}, B. \& {Battistini}, C. 2020, \aap, 633, A10

\bibitem[{{Blecic} {et~al.}(2016){Blecic}, {Harrington}, \&
  {Bowman}}]{Blecic2016}
{Blecic}, J., {Harrington}, J., \& {Bowman}, M.~O. 2016, \apjs, 225, 4

\bibitem[{{Boss}(1997)}]{Boss1997}
{Boss}, A.~P. 1997, Science, 276, 1836

\bibitem[{{Chase} {et~al.}(1982){Chase}, {Curnutt}, {Downey}, {McDonald},
  {Syverud}, \& {Valenzuela}}]{Chase1982}
{Chase}, M.~W., J., {Curnutt}, J.~L., {Downey}, J.~R., J., {et~al.} 1982,
  Journal of Physical and Chemical Reference Data, 11, 695

\bibitem[{{Chase}(1986)}]{Chase1986}
{Chase}, M.~W. 1986, {JANAF thermochemical tables}

\bibitem[{{Chen} {et~al.}(2002){Chen}, {Nissen}, {Zhao}, \&
  {Asplund}}]{Chen2002}
{Chen}, Y.~Q., {Nissen}, P.~E., {Zhao}, G., \& {Asplund}, M. 2002, \aap, 390,
  225

\bibitem[{{Ciesla}(2011)}]{Ciesla2011}
{Ciesla}, F.~J. 2011, \apj, 740, 9

\bibitem[{{D'Alessio} {et~al.}(1998){D'Alessio}, {Cant{\"o}}, {Calvet}, \&
  {Lizano}}]{DAlessio1998}
{D'Alessio}, P., {Cant{\"o}}, J., {Calvet}, N., \& {Lizano}, S. 1998, \apj,
  500, 411

\bibitem[{{D'Angelo} {et~al.}(2019){D'Angelo}, {Cazaux}, {Kamp}, {Thi}, \&
  {Woitke}}]{DAngelo2019}
{D'Angelo}, M., {Cazaux}, S., {Kamp}, I., {Thi}, W.~F., \& {Woitke}, P. 2019,
  \aap, 622, A208

\bibitem[{{Dorn} {et~al.}(2019){Dorn}, {Harrison}, {Bonsor}, \&
  {Hands}}]{2019ESS.....430006D}
{Dorn}, C., {Harrison}, J. H.~D., {Bonsor}, A., \& {Hands}, T. 2019, in
  AAS/Division for Extreme Solar Systems Abstracts, Vol.~51, AAS/Division for
  Extreme Solar Systems Abstracts, 300.06

\bibitem[{{Dullemond} {et~al.}(2007){Dullemond}, {Hollenbach}, {Kamp}, \&
  {D'Alessio}}]{Dullemond2007}
{Dullemond}, C.~P., {Hollenbach}, D., {Kamp}, I., \& {D'Alessio}, P. 2007, in
  Protostars and Planets V, ed. B.~{Reipurth}, D.~{Jewitt}, \& K.~{Keil}, 555

\bibitem[{{Dullemond} \& {Monnier}(2010)}]{Dullemond2010}
{Dullemond}, C.~P. \& {Monnier}, J.~D. 2010, \araa, 48, 205

\bibitem[{{Fu} \& {Elkins-Tanton}(2014)}]{Fu2014}
{Fu}, R.~R. \& {Elkins-Tanton}, L.~T. 2014, Earth and Planetary Science
  Letters, 390, 128

\bibitem[{{Gail} \& {Sedlmayr}(1986)}]{Gail1986}
{Gail}, H.~P. \& {Sedlmayr}, E. 1986, \aap, 166, 225

\bibitem[{{Hayashi}(1981)}]{Hayashi1981}
{Hayashi}, C. 1981, in IAU Symposium, Vol.~93, Fundamental Problems in the
  Theory of Stellar Evolution, ed. D.~{Sugimoto}, D.~Q. {Lamb}, \& D.~N.
  {Schramm}, 113--126

\bibitem[{{Herbort} {et~al.}(2021){Herbort}, {Woitke}, {Helling}, \&
  {Zerkle}}]{Herbort2021}
{Herbort}, O., {Woitke}, P., {Helling}, C., \& {Zerkle}, A. 2021, in EGU
  General Assembly Conference Abstracts, EGU General Assembly Conference
  Abstracts, EGU21--9745

\bibitem[{{Hinkel} {et~al.}(2014){Hinkel}, {Timmes}, {Young}, {Pagano}, \&
  {Turnbull}}]{Hinkel2014}
{Hinkel}, N.~R., {Timmes}, F.~X., {Young}, P.~A., {Pagano}, M.~D., \&
  {Turnbull}, M.~C. 2014, \aj, 148, 54

\bibitem[{{Hinkel} {et~al.}(2016){Hinkel}, {Young}, {Pagano}, {Desch}, {Anbar},
  {Adibekyan}, {Blanco-Cuaresma}, {Carlberg}, {Delgado Mena}, {Liu},
  {Nordlander}, {Sousa}, {Korn}, {Gruyters}, {Heiter}, {Jofr{\'e}}, {Santos},
  \& {Soubiran}}]{Hinkel2016}
{Hinkel}, N.~R., {Young}, P.~A., {Pagano}, M.~D., {et~al.} 2016, \apjs, 226, 4

\bibitem[{{Hirose} {et~al.}(2013){Hirose}, {Labrosse}, \&
  {Hernlund}}]{Hirose2013}
{Hirose}, K., {Labrosse}, S., \& {Hernlund}, J. 2013, Annual Review of Earth
  and Planetary Sciences, 41, 657

\bibitem[{{Johansen} {et~al.}(2021){Johansen}, {Ronnet}, {Bizzarro},
  {Schiller}, {Lambrechts}, {Nordlund}, \& {Lammer}}]{Johansen2021}
{Johansen}, A., {Ronnet}, T., {Bizzarro}, M., {et~al.} 2021, Science Advances,
  7, eabc0444

\bibitem[{{Johnson} {et~al.}(1992){Johnson}, {Oelkers}, \&
  {Helgeson}}]{Johnson1992}
{Johnson}, J.~W., {Oelkers}, E.~H., \& {Helgeson}, H.~C. 1992, Computers and
  Geosciences, 18, 899

\bibitem[{{Kobayashi} \& {Nakasato}(2011)}]{Kobayashi2011a}
{Kobayashi}, C. \& {Nakasato}, N. 2011, \apj, 729, 16

\bibitem[{{Kobayashi} {et~al.}(2011){Kobayashi}, {Kimura}, {Watanabe},
  {Yamamoto}, \& {M{\"u}ller}}]{Kobayashi2011b}
{Kobayashi}, H., {Kimura}, H., {Watanabe}, S.~i., {Yamamoto}, T., \&
  {M{\"u}ller}, S. 2011, Earth, Planets, and Space, 63, 1067

\bibitem[{{Kobayashi} {et~al.}(2009){Kobayashi}, {Watanabe}, {Kimura}, \&
  {Yamamoto}}]{Kobayashi2009}
{Kobayashi}, H., {Watanabe}, S.-i., {Kimura}, H., \& {Yamamoto}, T. 2009,
  \icarus, 201, 395

\bibitem[{{Laurenz} {et~al.}(2016){Laurenz}, {Rubie}, {Frost}, \&
  {Vogel}}]{laurenz2016}
{Laurenz}, V., {Rubie}, D.~C., {Frost}, D.~J., \& {Vogel}, A.~K. 2016, \gca,
  194, 123

\bibitem[{{Lewis}(1974)}]{lewis1974}
{Lewis}, J.~S. 1974, Science, 186, 440

\bibitem[{{Lichtenberg} {et~al.}(2021){Lichtenberg}, {Dr{\k{a}}{\.z}kowska},
  {Sch{\"o}nb{\"a}chler}, {Golabek}, \& {Hands}}]{Lichtenberg2021}
{Lichtenberg}, T., {Dr{\k{a}}{\.z}kowska}, J., {Sch{\"o}nb{\"a}chler}, M.,
  {Golabek}, G.~J., \& {Hands}, T.~O. 2021, Science, 371, 365

\bibitem[{{Liu} {et~al.}(2010){Liu}, {Sato}, {Takeda}, {Ando}, \&
  {Zhao}}]{Liu2010}
{Liu}, Y., {Sato}, B., {Takeda}, Y., {Ando}, H., \& {Zhao}, G. 2010, \pasj, 62,
  1071

\bibitem[{{Matteucci}(2021)}]{Matteucci2021}
{Matteucci}, F. 2021, \aapr, 29, 5

\bibitem[{{McDonough} \& {Sun}(1995)}]{McDonough1995}
{McDonough}, W.~F. \& {Sun}, S.~s. 1995, Chemical Geology, 120, 223

\bibitem[{{McDonough} \& {Yoshizaki}(2021)}]{mcdonough2021}
{McDonough}, W.~F. \& {Yoshizaki}, T. 2021, Progress in Earth and Planetary
  Science, 8, 39

\bibitem[{{Min} {et~al.}(2011){Min}, {Dullemond}, {Kama}, \&
  {Dominik}}]{Min2011}
{Min}, M., {Dullemond}, C.~P., {Kama}, M., \& {Dominik}, C. 2011, \icarus, 212,
  416

\bibitem[{{Noack} \& {Lasbleis}(2020)}]{Noack2020}
{Noack}, L. \& {Lasbleis}, M. 2020, \aap, 638, A129

\bibitem[{{Pollack} {et~al.}(1996){Pollack}, {Hubickyj}, {Bodenheimer},
  {Lissauer}, {Podolak}, \& {Greenzweig}}]{Pollack1996}
{Pollack}, J.~B., {Hubickyj}, O., {Bodenheimer}, P., {et~al.} 1996, \icarus,
  124, 62

\bibitem[{{Putirka} {et~al.}(2021){Putirka}, {Dorn}, {Hinkel}, \&
  {Unterborn}}]{Putirka2021}
{Putirka}, K., {Dorn}, C., {Hinkel}, N., \& {Unterborn}, C. 2021, arXiv
  e-prints, arXiv:2108.08383

\bibitem[{{Putirka} \& {Rarick}(2019)}]{Putirka2019}
{Putirka}, K.~D. \& {Rarick}, J.~C. 2019, American Mineralogist, 104, 817

\bibitem[{{Raymond} {et~al.}(2018){Raymond}, {Boulet}, {Izidoro}, {Esteves}, \&
  {Bitsch}}]{Raymond2018}
{Raymond}, S.~N., {Boulet}, T., {Izidoro}, A., {Esteves}, L., \& {Bitsch}, B.
  2018, \mnras, 479, L81

\bibitem[{{Raymond} \& {Morbidelli}(2020)}]{raymond2020}
{Raymond}, S.~N. \& {Morbidelli}, A. 2020, arXiv e-prints, arXiv:2002.05756

\bibitem[{{Schneider} \& {Bitsch}(2021{\natexlab{a}})}]{Schneider2021a}
{Schneider}, A.~D. \& {Bitsch}, B. 2021{\natexlab{a}}, \aap, 654, A71

\bibitem[{{Schneider} \& {Bitsch}(2021{\natexlab{b}})}]{Schneider2021b}
{Schneider}, A.~D. \& {Bitsch}, B. 2021{\natexlab{b}}, \aap, 654, A72

\bibitem[{{Schubert} {et~al.}(2001){Schubert}, {Turcotte}, \&
  {Olson}}]{Schubert2001}
{Schubert}, G., {Turcotte}, D.~L., \& {Olson}, P. 2001, {Mantle Convection in
  the Earth and Planets}

\bibitem[{{Spaargaren} {et~al.}(2020){Spaargaren}, {Ballmer}, {Bower}, {Dorn},
  \& {Tackley}}]{spaargaren2020}
{Spaargaren}, R.~J., {Ballmer}, M.~D., {Bower}, D.~J., {Dorn}, C., \&
  {Tackley}, P.~J. 2020, \aap, 643, A44

\bibitem[{{Thi} {et~al.}(2020){Thi}, {Hocuk}, {Kamp}, {Woitke}, {Rab},
  {Cazaux}, {Caselli}, \& {D'Angelo}}]{Thi2020}
{Thi}, W.~F., {Hocuk}, S., {Kamp}, I., {et~al.} 2020, \aap, 635, A16

\bibitem[{{Tuthill} {et~al.}(2001){Tuthill}, {Monnier}, \&
  {Danchi}}]{Tuthill2001}
{Tuthill}, P.~G., {Monnier}, J.~D., \& {Danchi}, W.~C. 2001, \nat, 409, 1012

\bibitem[{{Unterborn} {et~al.}(2016){Unterborn}, {Dismukes}, \&
  {Panero}}]{Unterborn2016}
{Unterborn}, C.~T., {Dismukes}, E.~E., \& {Panero}, W.~R. 2016, \apj, 819, 32

\bibitem[{{Urakawa} {et~al.}(2004){Urakawa}, {Someya}, {Terasaki}, {Katsura},
  {Yokoshi}, {Funakoshi}, {Utsumi}, {Katayama}, {Sueda}, \&
  {Irifune}}]{Urakawa2004}
{Urakawa}, S., {Someya}, K., {Terasaki}, H., {et~al.} 2004, Physics of the
  Earth and Planetary Interiors, 143, 469

\bibitem[{{Valencia} {et~al.}(2007){Valencia}, {Sasselov}, \&
  {O'Connell}}]{valencia2007}
{Valencia}, D., {Sasselov}, D.~D., \& {O'Connell}, R.~J. 2007, \apj, 665, 1413

\bibitem[{{Williams} \& {Cieza}(2011)}]{williams2011}
{Williams}, J.~P. \& {Cieza}, L.~A. 2011, \araa, 49, 67

\bibitem[{{Woitke} {et~al.}(2018){Woitke}, {Helling}, {Hunter}, {Millard},
  {Turner}, {Worters}, {Blecic}, \& {Stock}}]{Woitke2018}
{Woitke}, P., {Helling}, C., {Hunter}, G.~H., {et~al.} 2018, \aap, 614, A1

\bibitem[{{Zimmer} {et~al.}(2016){Zimmer}, {Zhang}, {Lu}, {Chen}, {Zhang},
  {Dalkilic}, \& {Zhu}}]{Zimmer2016}
{Zimmer}, K., {Zhang}, Y., {Lu}, P., {et~al.} 2016, Computers and Geosciences,
  90, 97

\end{thebibliography}

\begin{appendix}

\section{Initial elemental abundances (gas phase)}
\label{sec:chemical}

Table~\ref{fig:full chem} is a complete list of the initial elemental abundances in the gas phase for the 6 sample stars investigated in this paper (24 elements and 3 elemental ratios). The data has been taken from the Hypatia Catalog \citep{Hinkel2014}.

\begin{table}[h]

\caption{Initial elemental abundances in the gas phase}
\resizebox{\hsize}{!}{
\label{fig:full chem}
\begin{tabular}{llllllll}
\#    & 0   & 1                         & 2                        & 3                         & 4                         & 5                         & 6                         \\ \hline
HIP   & Sun & 109836                    & 28869                    & 63048                     & 43393                     & 99423                     & 83359                     \\ \hline
H     & 0.0 & 0.0                       & 0.0                      & 0.0                       & 0.0                       & 0.0                       & 0.0                       \\
He    & 0.0 & 0.0                       & 0.0                      & 0.0                       & 0.0                       & 0.0                       & 0.0                       \\
Fe    & 0.0 & +0.39                     & +0.13                    & -0.019                    & -0.57                     & -1.335                    & +0.145                    \\
C     & 0.0 & +0.225                    & +0.01                    & +0.13                     & -0.305                    & -0.49                     & -0.15                     \\
O     & 0.0 & +0.25                     & +0.02                    & +0.19                     & +0.03                     & -0.265                    & -0.03                     \\
Mg    & 0.0 & +0.15                     & -0.09                    & +0.25                     & -0.293                    & -0.93                     & +0.1                      \\
Si    & 0.0 & +0.285                    & +0.17                    & +0.08                     & -0.329                    & -0.92                     & +0.46                     \\
Ca    & 0.0 & +0.22                     & +0.1                     & +0.13                     & -0.44                     & -1.02                     & +0.26                     \\
Ti    & 0.0 & +0.285                    & +0.18                    & +0.11                     & -0.21                     & -1.065                    & +0.13                     \\
Li    & 0.0 & $\textcolor{red}{+0.277}$ & $\textcolor{red}{+0.12}$ & +0.06                     & $\textcolor{red}{-0.305}$ & +1.18                     & +0.145                    \\
N     & 0.0 & +0.29                     & $\textcolor{red}{+0.12}$ & $\textcolor{red}{+0.095}$ & $\textcolor{red}{-0.305}$ & +0.37                     & $\textcolor{red}{+0.185}$ \\
F     & 0.0 & $\textcolor{red}{+0.277}$ & $\textcolor{red}{+0.12}$ & $\textcolor{red}{+0.095}$ & $\textcolor{red}{-0.305}$ & $\textcolor{red}{-0.873}$ & $\textcolor{red}{+0.185}$ \\
Na    & 0.0 & +0.415                    & -0.02                    & +0.08                     & -0.46                     & -1.41                     & +0.46                     \\
Al    & 0.0 & +0.19                     & -0.08                    & +0.23                     & -0.263                    & -0.825                    & +0.38                     \\
P     & 0.0 & $\textcolor{red}{+0.277}$ & $\textcolor{red}{+0.12}$ & $\textcolor{red}{+0.095}$ & $\textcolor{red}{-0.305}$ & $\textcolor{red}{-0.873}$ & $\textcolor{red}{+0.185}$ \\
Cl    & 0.0 & $\textcolor{red}{+0.277}$ & $\textcolor{red}{+0.12}$ & $\textcolor{red}{+0.095}$ & $\textcolor{red}{-0.305}$ & $\textcolor{red}{-0.873}$ & $\textcolor{red}{+0.185}$ \\
K     & 0.0 & $\textcolor{red}{+0.277}$ & $\textcolor{red}{+0.12}$ & $\textcolor{red}{+0.095}$ & $\textcolor{red}{-0.305}$ & +0.36                     & $\textcolor{red}{+0.185}$ \\
V     & 0.0 & +0.25                     & +0.27                    & +0.06                     & -0.27                     & -0.95                     & +0.17                     \\
Cr    & 0.0 & +0.27                     & +0.21                    & +0.07                     & -0.49                     & -1.29                     & +0.26                     \\
Mn    & 0.0 & +0.33                     & $\textcolor{red}{+0.12}$ & +0.02                     & -0.73                     & -0.52                     & +0.2                      \\
Ni    & 0.0 & +0.38                     & +0.14                    & +0.036                    & -0.56                     & -1.34                     & +0.23                     \\
Zr    & 0.0 & $\textcolor{red}{+0.277}$ & $\textcolor{red}{+0.12}$ & +0.23                     & -0.2                      & +1.045                    & -0.16                     \\
W     & 0.0 & $\textcolor{red}{+0.277}$ & $\textcolor{red}{+0.12}$ & $\textcolor{red}{+0.095}$ & $\textcolor{red}{-0.305}$ & $\textcolor{red}{-0.873}$ & $\textcolor{red}{+0.185}$ \\
S     & 0.0 & $\textcolor{red}{+0.277}$ & +0.12                    & +0.23                     & +0.1                      & +0.21                     & +1.1                      \\ \hline
Mg/Si & 0.0 & -0.14                     & -0.26                    & +0.17                     & +0.04                     & -0.01                     & -0.36                     \\
Fe/Mg & 0.0 & +0.24                     & +0.22                    & -0.27                     & -0.28                     & -0.40                     & +0.04                     \\
Fe/S  & 0.0 & $\textcolor{red}{+0.11}$  & +0.01                    & -0.25                     & -0.67                     & -1.54                     & -0.96                    
\end{tabular}}
\tablefoot{The values are normalised with respect to the Solar abundance value \citep{Asplund2009} as described by Eq.~\ref{eq:4}. H and He are kept at the Solar abundance values for all stars. In black are the values from the Hypatia Catalog, while the red values are elements missing abundances in the database for which the values have been approximated using the median of the known data for the metals of the specific star.}

\end{table}

\section{Planet forming disk model}
\label{sec:App proto disk}

The inner radius $R_{\rm in}$ of the planet forming disk inside which dust grains cannot condense out from the gas is given by \citep{Tuthill2001},
\begin{equation}
R_{\rm in}=\frac{1}{2}\sqrt{Q_r}\bigg(\frac{T_*}{T_0}\bigg)^2R_*\hspace{10mm}\text{[cm]}
\end{equation}
with, $T_*$ the surface temperature of the host star, $R_*$ its radius, $T_0$ the sublimation temperature of the dust, and $Q_R$ the ratio of the absorption efficiencies at the given temperatures $T_*$ and $T_0$
\begin{equation}
Q_r=\frac{Q(T_*)}{Q(T_0)} \,\,\, .
\end{equation}
Here, we assume that the dust grains at the inner rim of the disk are radiating as a black body and hence $Q_R\!=\!1$. We assume a sublimation temperature $T_0\!=\!1500$~K \citep{Kobayashi2009, Kobayashi2011b}, above which most common dust species tend to sublimate. This is also consistent with interferometric observations that show the hottest dust thermal emission component (near-IR emission) to be at $\sim\!1500$~K \citep[see][]{Dullemond2010}.

\begin{figure}[htb]
    \centering
    \resizebox{\hsize}{!}{\includegraphics{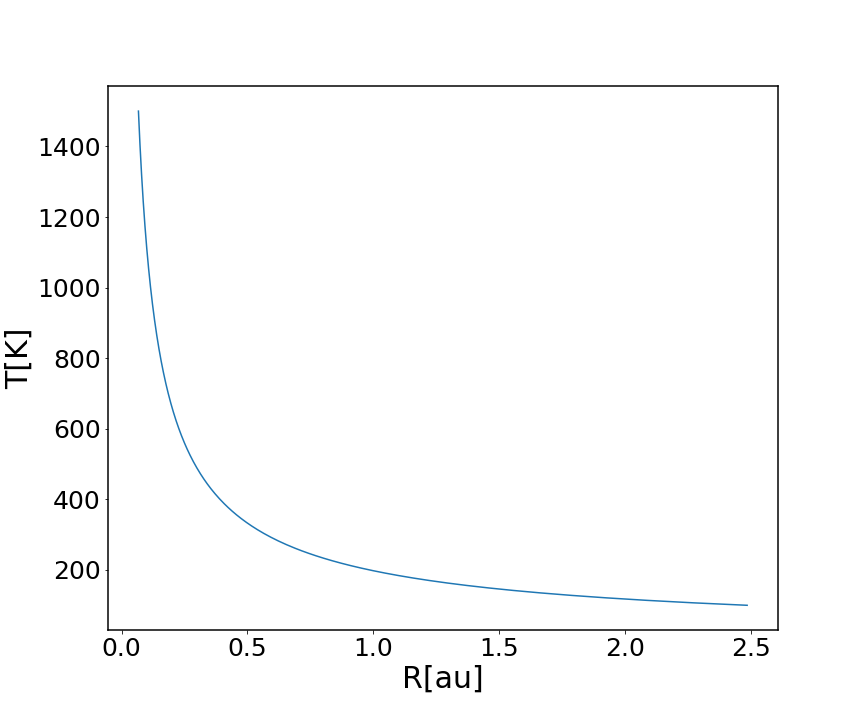}}
    \caption{Radial temperature profile in the mid-plane of the planet forming disk.}
    \label{fig:TR profile}
\end{figure}

Using the dust sublimation radius as the inner edge of the disk, we can calculate the midplane temperature $T$ as a function of the distance $r$ 
\begin{equation}
    T = T_0 \bigg(\frac{r}{R_{\rm in}}\bigg)^{-\frac{3}{4}} \hspace{10mm}[\text{K}]\label{eq:7} 
\end{equation}
following a simple power law approach which is supported by submm observations of planet forming disks \citep[][see Fig.~\ref{fig:TR profile})]{williams2011}. 

The surface density of the disk can be parametrized using the following equation
\begin{equation}
    \Sigma=\Sigma_0\bigg(\frac{r}{{\rm AU}}\bigg)^{-\frac{3}{2}}\hspace{10mm}\text{[g cm$^{-2}$]} \label{eq:8}
\end{equation}
with the Minimum Mass Solar Nebula\footnote{The Minimum Mass Solar Nebula represents the minimum amount of nebula material that must have been present to form the planets of the Solar system as we know it.} $\Sigma_0\!=\!1700$~g$/$cm$^{2}$ as the normalisation constant \citep{Armitage2019,Hayashi1981}. 

To obtain the midplane density and hence pressure, we need to calculate the gas scale height 
\begin{equation}
    H=\frac{c_{\rm s}}{\Omega}\hspace{10mm}\text{[cm]} \label{eq:9}
\end{equation}
with, the sound speed $c_{\rm s}$ and the Keplerian angular frequency $\Omega$ \citep{DAlessio1998}. The sound speed can be calculated under the assumption that the disk is vertically isothermal
\begin{equation}
    c_{\rm s}=\sqrt{\frac{k_{\rm B} T}{\mu m_{\rm p}}}\hspace{10mm}\text{[cm s$^{-1}$]}
\end{equation}
with, $k_{\rm B}$ the Boltzmann constant, $\mu\!=\!2.3$ the mean molecular weight, and $m_{\rm p}$ the mass of a proton. The Keplerian angular frequency is given by,
\begin{equation}
    \Omega=\sqrt{\frac{GM_*}{r^3}}\hspace{10mm}\text{[s$^{-1}$]}
\end{equation}
with, $G$ the gravitational constant and $M_*$ the mass of the host star.

Assuming that the ideal gas law holds, and combining Eq.~\ref{eq:8}, and Eq.~\ref{eq:9} the density and the pressure can be calculated as follows,
\begin{equation}
    \rho=\frac{1}{2}\pi\frac{\Sigma}{H}\hspace{10mm}\text{[g~cm$^{-3}$]}
\end{equation}
\begin{equation}
    p=\rho\frac{R_{\rm c}}{\mu N_0 m_{\rm p}}T\hspace{10mm}\text{[dyn~cm$^{-2}$]}
\end{equation}
with, Avogadro's constant $N_0$ in mol$^{-1}$, and the ideal gas constant $R_{\rm c}$ in erg~ (K~mol)$^{-1}$. 

\section{Complete data from condensation sequences of the sample of stars}
\label{sec:condensate}

\begin{figure}[thb]
    \centering
    
    \includegraphics[width=8.5cm]{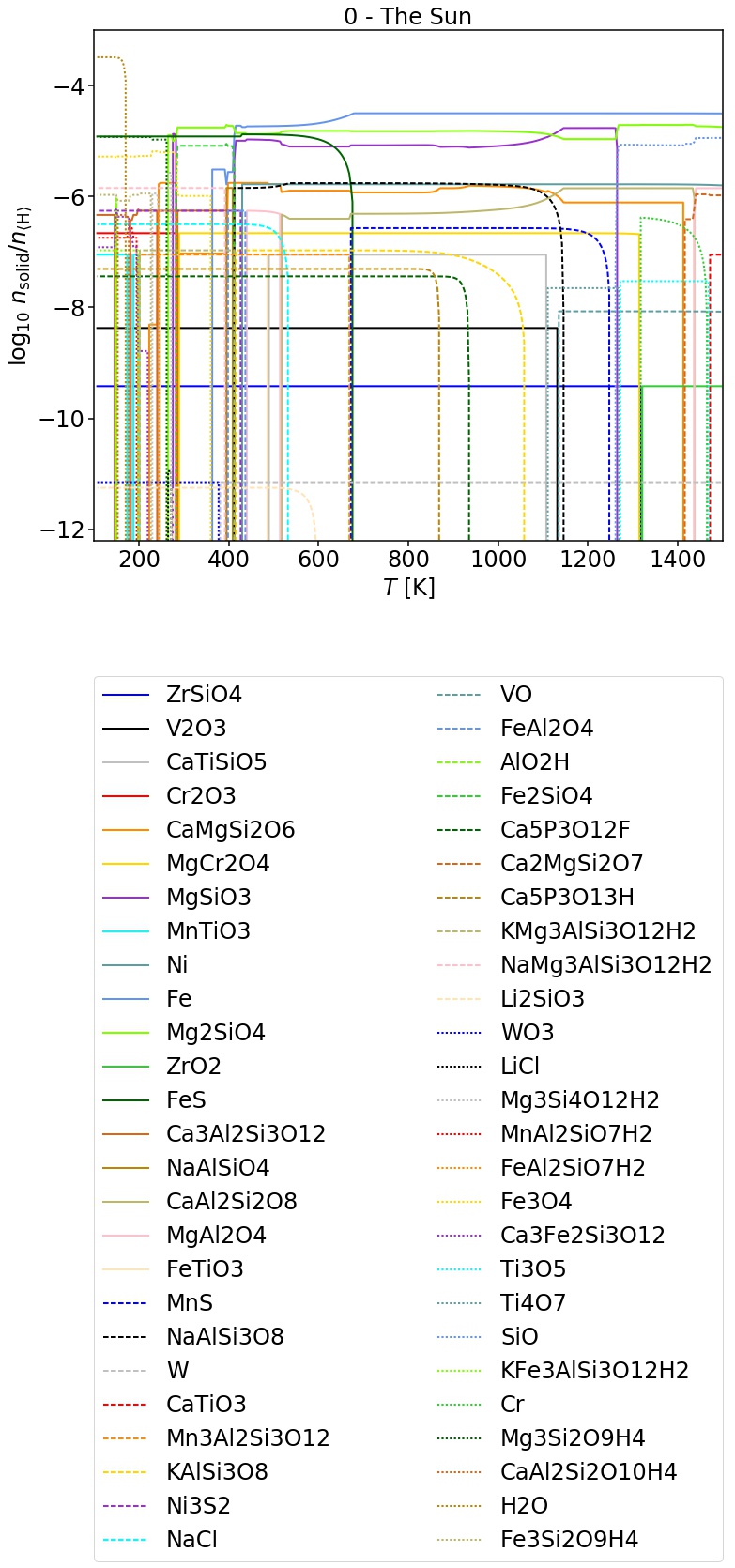}
    
    \caption{Condensation sequence for the Sun at non constant pressure (see Fig.~\ref{fig:pT profile}). Smooth condensation (round edge) represents phase transition gas/solid, while sharp condensation represents phase transition solid/solid.}
    \label{fig:Cond sol}
\end{figure}

\begin{figure}[thb]
    \centering
    \includegraphics[width=8.5cm]{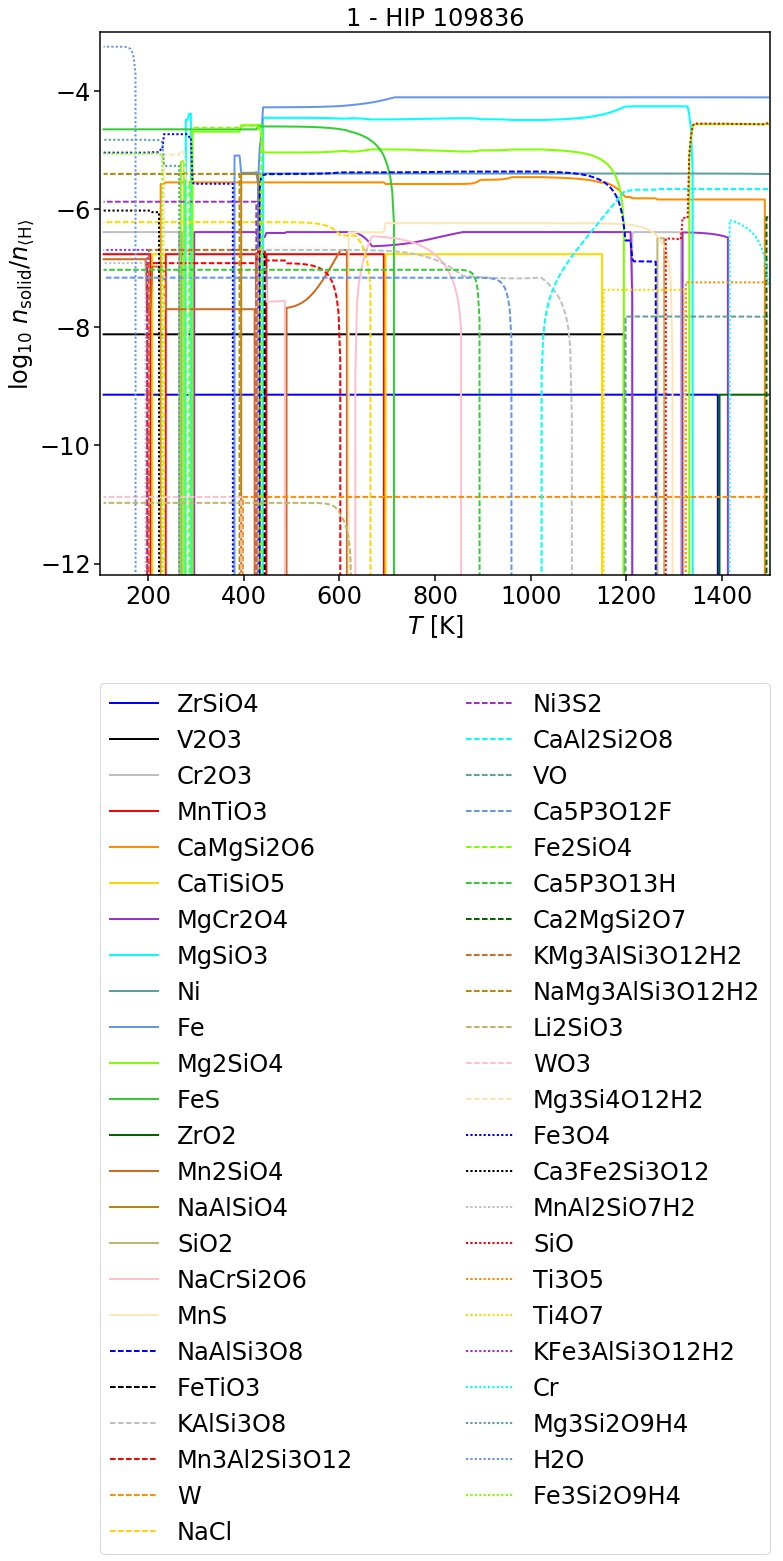}
    \caption{Same as Fig.~\ref{fig:Cond sol} for HIP 109836 (star 1).}
    \label{fig:Cond 109836}
\end{figure}

\begin{figure}[thb]
    \centering
    \includegraphics[width=8.5cm]{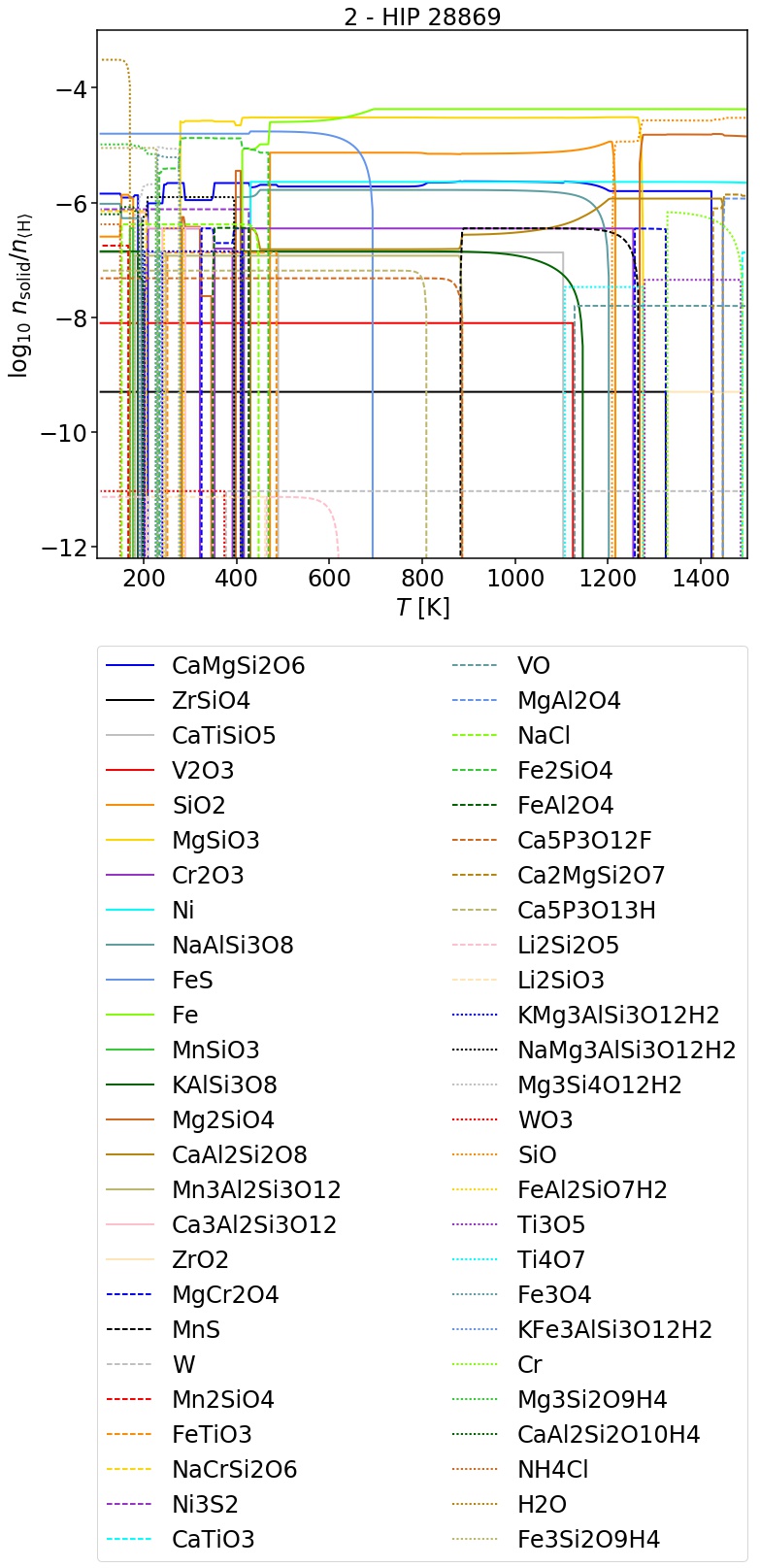}
    \caption{Same as Fig.~\ref{fig:Cond sol} for HIP 28869 (star 2)}
    \label{fig:Cond 28869}
\end{figure}

\begin{figure}[thb]
    \centering
    \includegraphics[width=8.5cm]{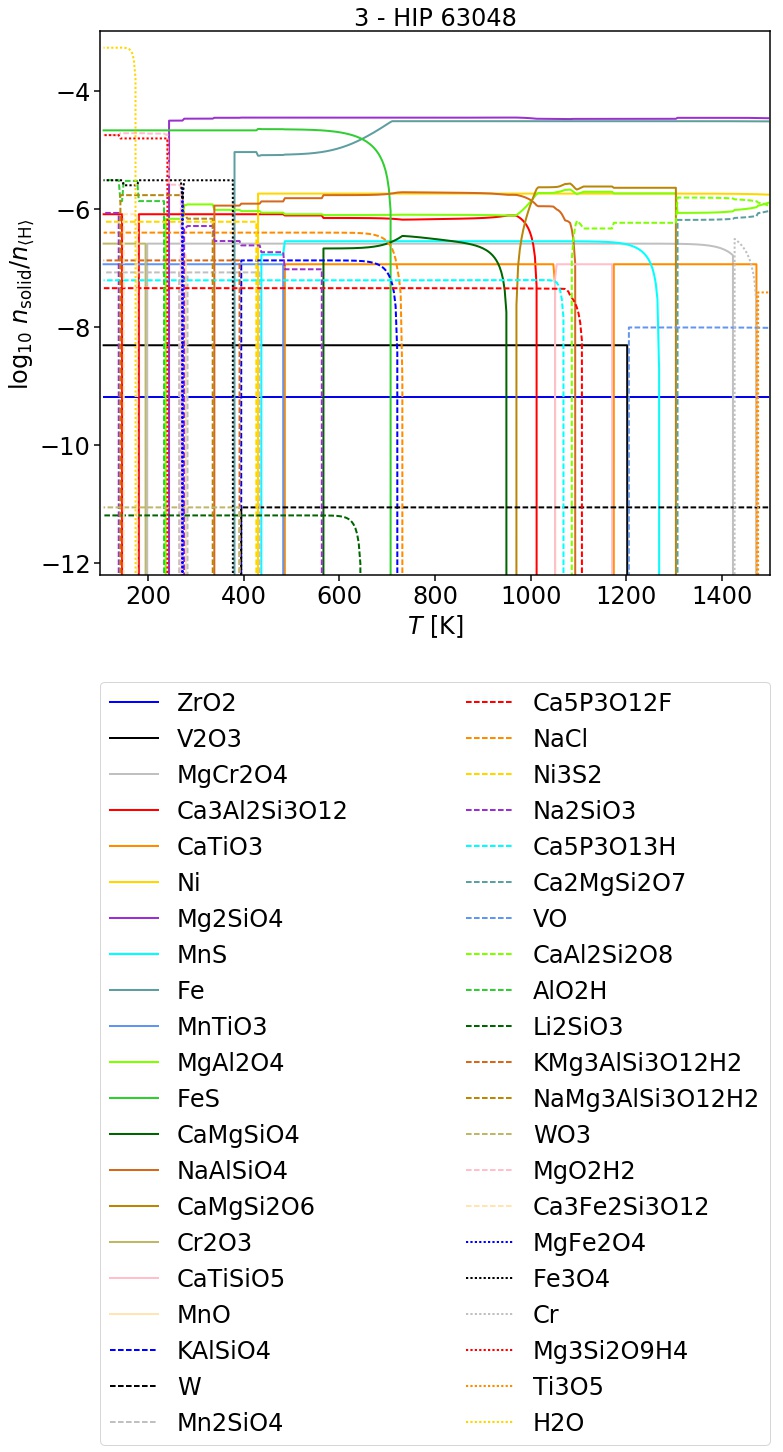}
    \caption{Same as Fig.~\ref{fig:Cond sol} for HIP 63048 (star 3)}
    \label{fig:Cond 63048}
\end{figure}

\begin{figure}[thb]
    \centering
    \includegraphics[width=8.5cm]{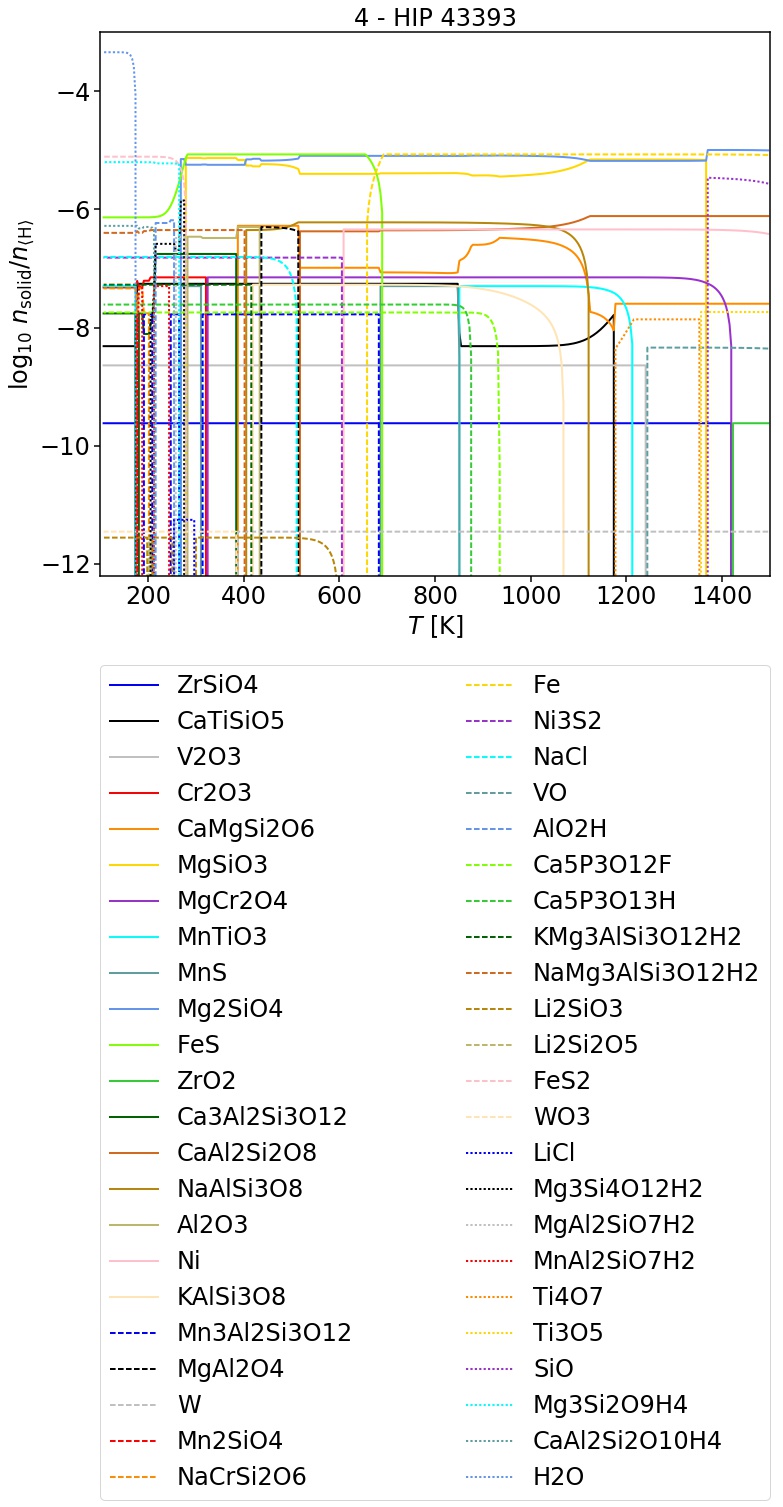}
    \caption{Same as Fig.~\ref{fig:Cond sol} for HIP 43393 (star 4)}
    \label{fig:Cond 43393}
\end{figure}

\begin{figure}[thb]
    \centering
    \includegraphics[width=8.5cm]{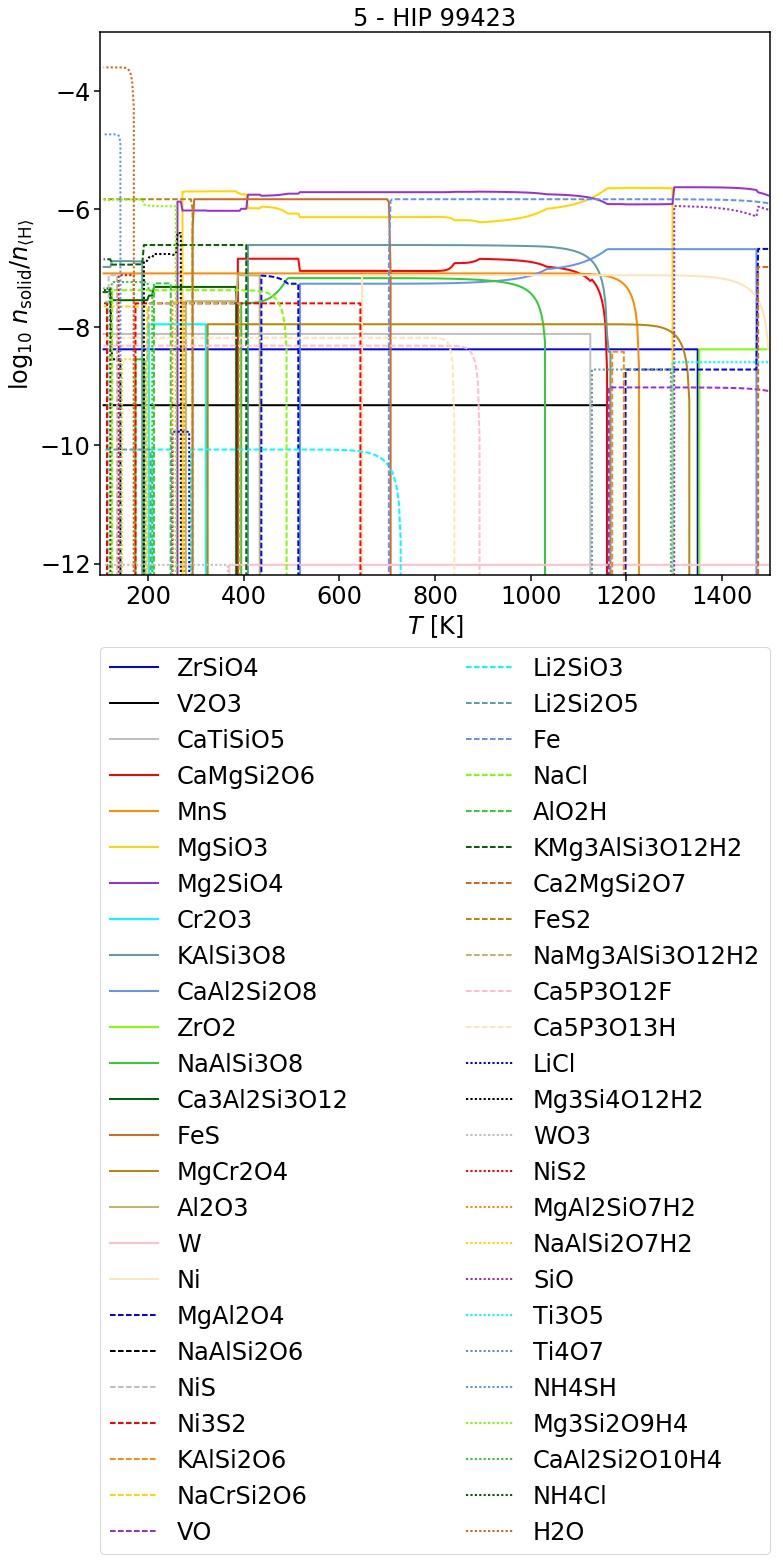}
    \caption{Same as Fig.~\ref{fig:Cond sol} for HIP 99423 (star 5)}
    \label{fig:Cond 99423}
\end{figure}

\begin{figure}[thb]
    \centering
    \includegraphics[width=8.5cm]{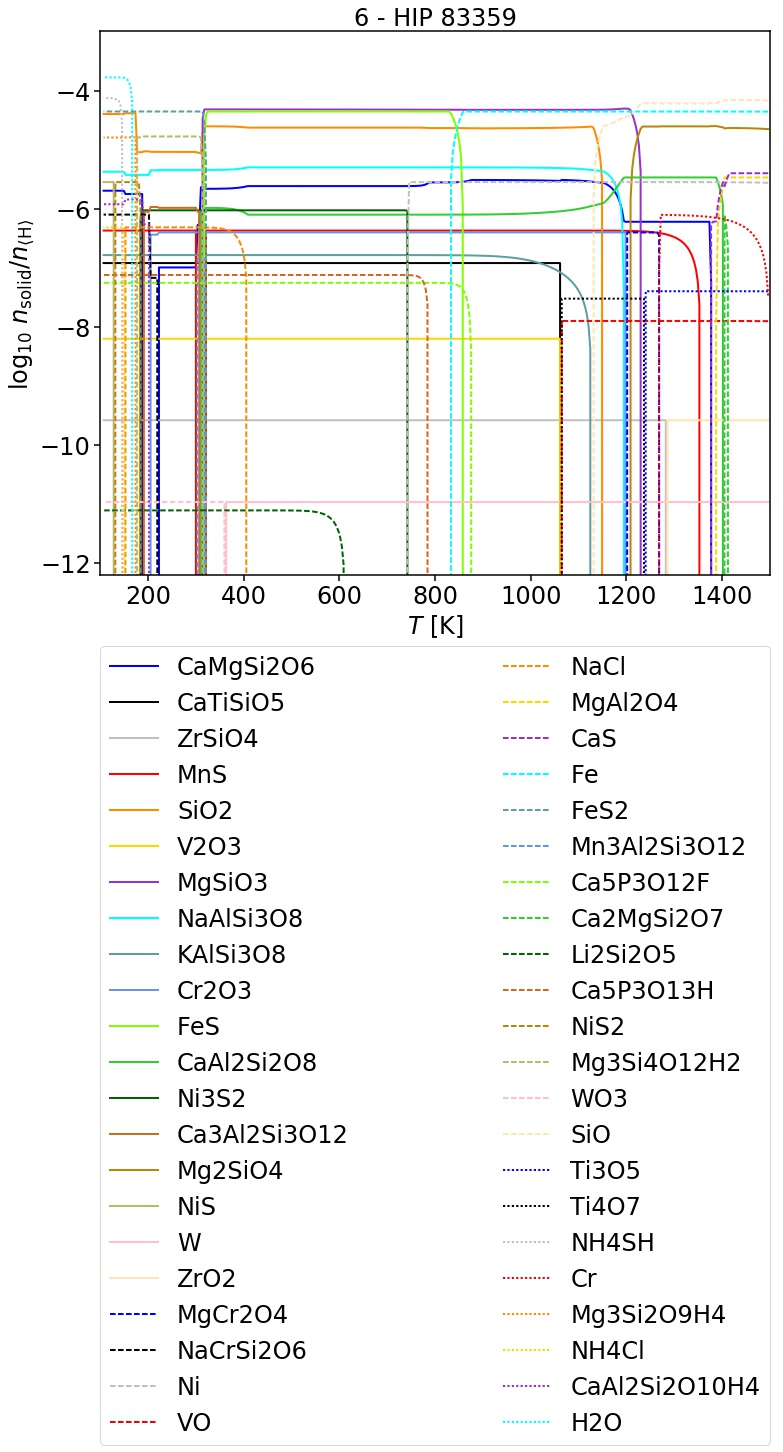}
    \caption{Same as Fig.~\ref{fig:Cond sol} for HIP 83359 (star 6)}
    \label{fig:Cond 83359}
\end{figure}

We present here the full condensation sequences for all condensates forming in our disk model around the 6 sample stars (Figs.~\ref{fig:Cond sol} to \ref{fig:Cond 83359}). Tables~\ref{tab:app-1} and \ref{tab:app-2} represent all mass fractions and elemental ratios of the condensed phase that are used throughout this paper.


\begin{table}[th]
\caption{Mass fraction of the hypothetical planets forming in situ. Values are given as a percentage of the total mass of the body [wt$\%$].}
\begin{tabular}{lllllll}
Planet & Star & O     & Mg   & Si   & Fe   & S     \\ \hline
1500 K & 0    & 28.66 & 17.30 & 16.42 & 32.74 & 0.00  \\
       & 1    & 24.56 & 12.92 & 16.45 & 41.28 & 0.00  \\
       & 2    & 25.41 & 11.80 & 20.65 & 37.25 & 0.00  \\
       & 3    & 34.26 & 23.32 & 14.61 & 22.96 & 0.00  \\
       & 4    & 35.11 & 21.73 & 17.94 & 20.79 & 0.00  \\
       & 5    & 36.31 & 21.77 & 19.62 & 17.19 & 0.00  \\
       & 6    & 28.65 & 12.20 & 26.58 & 25.86 & 1.35  \\ \hline
1250 K & 0    & 31.27 & 16.92 & 15.89 & 30.88 & 0.00  \\
       & 1    & 27.72 & 12.29 & 15.74 & 38.96 & 0.12  \\
       & 2    & 28.93 & 11.46 & 19.59 & 34.71 & 0.07  \\
       & 3    & 34.28 & 23.08 & 14.66 & 22.67 & 0.05  \\
       & 4    & 36.49 & 20.80 & 17.98 & 20.06 & 0.00  \\
       & 5    & 39.25 & 20.89 & 20.07 & 15.00 & 0.00  \\
       & 6    & 30.76 & 12.23 & 26.32 & 24.76 & 0.13  \\ \hline
1000 K & 0    & 31.09 & 16.70 & 15.68 & 30.47 & 0.15  \\
       & 1    & 27.67 & 12.13 & 15.54 & 38.46 & 0.16  \\
       & 2    & 30.68 & 11.04 & 18.87 & 33.44 & 0.16  \\
       & 3    & 34.23 & 22.83 & 14.50 & 22.43 & 0.12  \\
       & 4    & 36.39 & 20.61 & 17.82 & 19.88 & 0.00  \\
       & 5    & 38.30 & 20.17 & 19.38 & 14.48 & 0.46  \\
       & 6    & 36.76 & 10.98 & 23.62 & 22.23 & 0.12  \\ \hline
401 K  & 0    & 32.00 & 14.84 & 13.94 & 27.09 & 6.48  \\
       & 1    & 30.64 & 10.55 & 15.53 & 33.47 & 6.18  \\
       & 2    & 32.33 & 9.66  & 16.50 & 29.25 & 6.84  \\
       & 3    & 31.22 & 20.76 & 13.18 & 20.39 & 8.66  \\
       & 4    & 32.70 & 18.26 & 15.78 & 17.61 & 10.47 \\
       & 5    & 35.57 & 18.28 & 17.57 & 13.13 & 8.22  \\
       & 6    & 32.47 & 9.66  & 20.79 & 19.56 & 11.82 \\ \hline
253 K  & 0    & 36.81 & 13.65 & 12.82 & 24.90 & 5.96  \\
       & 1    & 34.22 & 9.96  & 12.76 & 31.58 & 5.83  \\
       & 2    & 34.36 & 9.33  & 15.95 & 28.26 & 6.61  \\
       & 3    & 33.31 & 20.10 & 12.76 & 19.74 & 8.39  \\
       & 4    & 34.43 & 15.65 & 13.53 & 15.09 & 16.05 \\
       & 5    & 36.52 & 15.95 & 15.33 & 11.46 & 13.75 \\
       & 6    & 30.40 & 8.50  & 18.29 & 17.20 & 20.27 \\ \hline
198 K  & 0    & 37.14 & 13.55 & 12.73 & 24.74 & 5.92  \\
       & 1    & 35.52 & 9.70  & 12.43 & 30.76 & 5.68  \\
       & 2    & 36.58 & 8.93  & 15.26 & 27.04 & 6.33  \\
       & 3    & 38.37 & 18.26 & 11.60 & 17.94 & 7.62  \\
       & 4    & 34.43 & 15.43 & 13.34 & 14.89 & 16.67 \\
       & 5    & 36.74 & 15.88 & 15.27 & 11.41 & 13.69 \\
       & 6    & 30.48 & 8.49  & 18.26 & 17.18 & 20.25 \\ \hline
144 K  & 0    & 60.35 & 7.47  & 7.01  & 13.63 & 3.26  \\
       & 1    & 57.69 & 5.66  & 7.26  & 17.97 & 3.32  \\
       & 2    & 56.84 & 5.46  & 9.33  & 16.53 & 3.87  \\
       & 3    & 63.80 & 9.06  & 5.75  & 8.90  & 3.78  \\
       & 4    & 73.57 & 4.33  & 3.74  & 4.17  & 4.67  \\
       & 5    & 81.59 & 2.19  & 2.10  & 1.57  & 1.95  \\
       & 6    & 38.10 & 6.36  & 13.68 & 12.87 & 20.33 \\ \hline
100 K  & 0    & 60.38 & 7.46  & 7.00  & 13.61 & 3.26  \\
       & 1    & 57.71 & 5.66  & 7.25  & 17.95 & 3.31  \\
       & 2    & 56.88 & 5.45  & 9.32  & 16.51 & 3.86  \\
       & 3    & 63.82 & 9.05  & 5.75  & 8.89  & 3.78  \\
       & 4    & 73.59 & 4.32  & 3.73  & 4.17  & 4.66  \\
       & 5    & 69.22 & 1.85  & 1.78  & 1.33  & 11.15 \\
       & 6    & 34.01 & 5.67  & 12.19 & 11.47 & 24.75
\end{tabular}
\label{tab:app-1}
\end{table}

\begin{table}[th]
\caption{Element ratios found in the condensed phase from which the hypothetical planets form in situ. Ratios are normalised to the Solar value (Eq.~\ref{eq:4}).}
\begin{tabular}{lllllll}
Planet & Star & Si/O  & Fe/O  & Fe/Si & Mg/Si & Fe/S  \\ \hline
1500 K & 0    & 0.70  & 0.70  & 0.01  & 0.00  & 294   \\
       & 1    & 0.76  & 0.87  & 0.11  & -0.13 & 294   \\
       & 2    & 0.85  & 0.81  & -0.03 & -0.27 & 294   \\
       & 3    & 0.57  & 0.47  & -0.09 & 0.18  & 294   \\
       & 4    & 0.64  & 0.42  & -0.22 & 0.06  & 293   \\
       & 5    & 0.67  & 0.32  & -0.35 & 0.02  & 292   \\
       & 6    & 0.90  & 0.60  & -0.30 & -0.37 & 0.66  \\ \hline
1250 K & 0    & 0.64  & 0.64  & 0.00  & 0.00  & 294   \\
       & 1    & 0.69  & 0.79  & 0.11  & -0.13 & 1.88  \\
       & 2    & 0.77  & 0.72  & -0.04 & -0.26 & 2.10  \\
       & 3    & 0.57  & 0.46  & -0.10 & 0.17  & 2.03  \\
       & 4    & 0.63  & 0.38  & -0.24 & 0.04  & 293   \\
       & 5    & 0.64  & 0.23  & -0.41 & -0.01 & 292   \\
       & 6    & 0.87  & 0.55  & -0.31 & -0.36 & 1.66  \\ \hline
1000 K & 0    & 0.64  & 0.64  & 0.00  & 0.00  & 1.69  \\
       & 1    & 0.69  & 0.79  & 0.11  & -0.13 & 1.75  \\
       & 2    & 0.73  & 0.68  & -0.04 & -0.26 & 1.70  \\
       & 3    & 0.56  & 0.46  & -0.10 & 0.17  & 1.65  \\
       & 4    & 0.63  & 0.38  & -0.24 & 0.04  & 293   \\
       & 5    & 0.64  & 0.22  & -0.41 & -0.01 & 0.87  \\
       & 6    & 0.74  & 0.43  & -0.31 & -0.36 & 1.63  \\ \hline
401 K  & 0    & 0.58  & 0.57  & 0.00  & 0.00  & 0.00  \\
       & 1    & 0.58  & 0.68  & 0.11  & -0.13 & 0.11  \\
       & 2    & 0.64  & 0.60  & -0.04 & -0.26 & 0.01  \\
       & 3    & 0.56  & 0.46  & -0.10 & 0.17  & -0.25 \\
       & 4    & 0.62  & 0.38  & -0.24 & 0.04  & -0.40 \\
       & 5    & 0.63  & 0.21  & -0.41 & -0.01 & -0.42 \\
       & 6    & 0.74  & 0.42  & -0.31 & -0.36 & -0.40 \\ \hline
253 K  & 0    & 0.48  & 0.47  & 0.00  & 0.00  & 0.00  \\
       & 1    & 0.51  & 0.61  & 0.11  & -0.13 & 0.13  \\
       & 2    & 0.60  & 0.56  & -0.04 & -0.26 & 0.01  \\
       & 3    & 0.52  & 0.42  & -0.10 & 0.17  & -0.25 \\
       & 4    & 0.53  & 0.29  & -0.24 & 0.04  & -0.65 \\
       & 5    & 0.56  & 0.14  & -0.41 & -0.01 & -0.70 \\
       & 6    & 0.72  & 0.40  & -0.31 & -0.36 & -0.69 \\ \hline
198 K  & 0    & 0.47  & 0.47  & 0.00  & 0.00  & 0.00  \\
       & 1    & 0.48  & 0.58  & 0.11  & -0.13 & 0.11  \\
       & 2    & 0.56  & 0.51  & -0.04 & -0.26 & 0.01  \\
       & 3    & 0.42  & 0.31  & -0.10 & 0.17  & -0.25 \\
       & 4    & 0.52  & 0.28  & -0.24 & 0.04  & -0.67 \\
       & 5    & 0.55  & 0.14  & -0.41 & -0.01 & -0.70 \\
       & 6    & 0.71  & 0.40  & -0.31 & -0.36 & -0.69 \\ \hline
144 K  & 0    & 0.00  & 0.00  & 0.00  & 0.00  & 0.00  \\
       & 1    & 0.04  & 0.14  & 0.11  & -0.13 & 0.11  \\
       & 2    & 0.15  & 0.11  & -0.04 & -0.26 & 0.01  \\
       & 3    & -0.11 & -0.21 & -0.10 & 0.17  & -0.25 \\
       & 4    & -0.36 & -0.60 & -0.24 & 0.04  & -0.67 \\
       & 5    & -0.65 & -1.07 & -0.41 & -0.01 & -0.71 \\
       & 6    & 0.49  & 0.17  & -0.31 & -0.36 & -0.82 \\ \hline
100 K  & 0    & 0.00  & 0.00  & 0.00  & 0.00  & 0.00  \\
       & 1    & 0.04  & 0.14  & 0.11  & -0.13 & 0.11  \\
       & 2    & 0.15  & 0.11  & -0.04 & -0.26 & 0.01  \\
       & 3    & -0.11 & -0.21 & -0.10 & 0.17  & -0.25 \\
       & 4    & -0.36 & -0.60 & -0.24 & 0.04  & -0.67 \\
       & 5    & -0.65 & -1.07 & -0.41 & -0.01 & -1.55 \\
       & 6    & 0.49  & 0.17  & -0.31 & -0.36 & -0.96
\end{tabular}
\label{tab:app-2}
\end{table}

\end{appendix}

\end{document}